\documentclass[reprint,showpacs,amsmath,amssymb,aps,pra,superscriptaddress,nobibnotes]{revtex4-2}

\usepackage{graphicx}
\usepackage{dcolumn}
\usepackage{bm}
\usepackage{enumerate}
\usepackage[ruled]{algorithm2e}
\usepackage{tabularx}
\usepackage{placeins}
\usepackage{threeparttable}
\usepackage[colorlinks, linkcolor=red, anchorcolor=blue, citecolor=green]{hyperref}
\usepackage[caption=false]{subfig}
\DeclareMathOperator{\Tr}{Tr}

\begin{document}

\title{Quantum optimal control of multi-level dissipative quantum systems with Reinforcement Learning}

\author{Zheng An}  \author{Hai-Jing Song}

\affiliation{Institute of Physics, Beijing National Laboratory for
  Condensed Matter Physics,\\Chinese Academy of Sciences, Beijing
  100190, China}

\affiliation{School of Physical Sciences, University of Chinese
  Academy of Sciences, Beijing 100049, China}

\author{Qi-Kai He}

\affiliation{HangZhou Tuya Information Technology Co., Ltd, Hangzhou, Zhejiang 310000, China}

\author{D. L. Zhou} \email[]{zhoudl72@iphy.ac.cn}

\affiliation{Institute of Physics, Beijing National Laboratory for
  Condensed Matter Physics,\\Chinese Academy of Sciences, Beijing
  100190, China}

\affiliation{School of Physical Sciences, University of Chinese
  Academy of Sciences, Beijing 100049, China}
\affiliation{Collaborative Innovation Center of Quantum Matter,
  Beijing 100190, China}

\affiliation{Songshan Lake Materials Laboratory, Dongguan, Guangdong
  523808, China}

\date{\today}

\begin{abstract}
  Manipulate and control of the complex quantum system with high
  precision are essential for achieving universal fault tolerant
  quantum computing. For a physical system with restricted control
  resources, it is a challenge to control the dynamics of the target
  system efficiently and precisely under disturbances. Here we propose
  a multi-level dissipative quantum control framework and show that deep
  reinforcement learning provides an efficient way to identify the
  optimal strategies with restricted control parameters of the complex
  quantum system. This framework can be generalized to be applied to
  other quantum control models. Compared with the traditional optimal
  control method, this deep reinforcement learning algorithm can
  realize efficient and precise control for multi-level quantum
  systems with different types of disturbances.
\end{abstract}


\maketitle

\section{Introduction \label{sec:intro}}

Precise and complete control of complex quantum systems is the core to
achieve quantum computation and quantum information processing. The
quantum control (QC) theory provides a powerful tool to achieve high
precision control of quantum dynamics. A QC problem can be phrased as
finding strategies of inducing complete transfer of population from an
arbitrary initial quantum state to the desired target state. An
optimal strategy to get a selected state of a finite energy level
quantum system is of primary importance for the control of quantum
dynamics. The theory for design such an optimal strategy has been
studied widely, such as Lyapunov quantum
control~\cite{lyapunov1,lyapunov2}, geometric control
theory~\cite{geo}, and Pontryagin maximum principle~\cite{pontryagin}.
Also, robust and optimal strategies of QC is essential for many areas
of physical systems from nitrogen-vacancy center
experiments~\cite{NVC}, optical systems\cite{optical} to
superconducting qubits~\cite{superconducting}. However, it is hard to
get a convincing result with traditional control theory if there have
some restricted conditions in the control system. To manipulate
more complicated systems, there have been developed sever algorithms
in numerical, like GRAPE~\cite{grape,grape2} and
CRAB~\cite{crab1,crab2}. Further, the disturbance of quantum dynamics is
the main obstacle in implementing scalable quantum
computing~\cite{nielsen}. To deal with the spin or qubit decoherence,
various strategies have been developed, including quantum error
correction~\cite{QEC1,QEC2,QEC3,QEC4}, dynamical decoupling
(DD)~\cite{DD1,DD2,DD3}, and optimized control in protecting quantum
coherence~\cite{ds1,ds2,Palittapongarnpim2017}. One way to
achieve the optimal control is to use an arbitrarily slow change of
the dynamical parameters and the adiabatic theorem~\cite{born1928}.
However, for a multi-level system, these require several resources
that also increase exponentially with the size of the system. On the other hand, when applied to a typical realistic
condition of an open quantum system is considered, there are few analytical or ansatz
solutions available. To
simplify those constraints, here we introduce a switch on-off control
problem with dissipative dynamics in this paper. In particular, we discuss
the dynamics that are affected by
dephasing and energy decay. These two effects exist, to
different degrees, in any practical attempt to implement quantum control tasks in real physical systems~\cite{noise1,noise2,noise3,noise4}. Those disturbances effects emerge from
the interaction of the system with the surrounding environment~\cite{open}.

Quantum control theory has been recently applied with success
to the optimization of the dynamics of simple systems~\cite{optimal1,optimal2,optimal3,optimal4} and quantum many body systems~\cite{crab1,crab2,Bukov2017,PhysRevA.84.012312}.
With the progress of quantum control techniques and computer science, the
numerical algorithm gives us a robust and efficient way to implement
high-fidelity quantum control. Among various control algorithms,
reinforcement learning (RL) has been attracting much focus. Reinforcement
learning has demonstrated remarkable abilities in board
games~\cite{alphago,alphagozero,alphazero} and video
games~\cite{atari,alphastar,multi}. Recently it has also been widely
applied to a wide array of physics problems, such as quantum state
preparation~\cite{Bukov2017,rlqc}, quantum gate
control~\cite{gate,Niu2019}, quantum error correction~\cite{error},
and quantum metrology~\cite{Xu2019}. Those successes naturally raise the
question of how much quantum control might benefit from the
application of reinforcement learning.

In this paper,
we study a general quantum control model of a finite-level system under disturbances. To explore the
optimal strategy of the control problem in this scenario, we use the
distributed proximal policy optimization (DPPO)
algorithm~\cite{DPPO1,DPPO2} to study this problem in this paper. The
proximal policy optimization (PPO) algorithm has been successfully
used in robotics~\cite{Tieck2018} and aircraft
control~\cite{aircraft}. Recently, it has been applied in QC problems
~\cite{DBLP,chen2019manipulation}.

The rest of this paper is structured as follows. In Sec.~\ref{model1},
we briefly introduce the basic description of our quantum control
model. In Sec.~\ref{AC}, we present the Actor-Critic model of
Reinforcement learning and DPPO algorithm used in our paper. In
Sec.~\ref{sec:quant-state-contr}, we present the methodology of
our method, the architecture of the neural network for our agent, the
interactive interface as well as numerical results of tested examples.
Finally, in Sec.~\ref{con}, we draw our conclusions.

\section{Model \label{model1}}

We study a quantum system with a finite number of distinct energy
levels driven by a time-dependent external field whose Hamiltonian
reads:
\begin{equation}
  H = H_0 + V
\end{equation}
with
\begin{align}
  \label{eq:1}
  H_{0} & = \sum_{i=1}^{n} E_{i} |i\rangle  \langle i|, \\
  V(t) & = \sum_{i=1}^{n-1} \gamma(t) \left( |i\rangle  \langle i+1| + |i\rangle  \langle
         i+1| \right),\label{eq:2}
\end{align}
where $H_0$ is called the drift Hamiltonian and $V(t)$ is called the
control Hamiltonian in quantum control theory. The state $|i\rangle$
is the $i$-th eigenstate of $H_{0}$ with eigenenergy $E_{i}$, $n$ is
the number of the energy levels, and the time-dependent real parameter
$\gamma(t)$ is the coupling strength between $|i\rangle$ and
$|i+1\rangle$ for $1\le i\le n-1$. Without losing of generality, we
assume that $E_{1}\le E_{2}\le \cdots\le E_{n}$. In particular, we assume $H_0$ is regular, where the energy levels
$E_i=i\ (i=1,\dots,n)$. However, a different distribution of eigenenergies may affect 
the performance of control algorithms. So we presented the effect of the different distribution
of eigenenergies on two examples in Appendix~\ref{eng}.

When our system weakly interacts with its environment, its dynamics is
described by the master equation of the Lindblad type:
\begin{equation}
  \label{eq:evo}
  \dot{\rho} = - \frac{i}{\hbar} [H, \rho] + \sum_{k} \Gamma_{k,n} \left(A_{k,n} \rho A_{k,n}^{\dagger}
    - \frac{1}{2} \left\{A_{k,n}^{\dagger} A_{k,n}, \rho\right\} \right),    
\end{equation}
where $A_{k,n}$ is the Lindblad operator associated with some dissipative
process with a decay rate $\Gamma_{k,n}$ for each $k$, and the subscript $n$
labels the type of dissipative process.
$\{A, B\} = AB + BA$ denotes the anticommutator. Here we consider two
typical dissipative processes. One is the dephasing process, whose
Lindblad operator $A_{k,d}=|k\rangle\langle k|$ with an identical
dephasing rate $\Gamma_{k,d}=\Gamma_{d}$ for $1\le k\le n$. The other is
the energy decay process, whose Lindblad operator
$A_{k,l}=|1\rangle\langle k|$ with an identical energy decay rate
$\Gamma_{k,l}=\Gamma_{l}$ for $2\le k\le n$.

Our central task can be stated as follows. Initially, our system is
prepared in the ground state $|1\rangle $ of $H_{0}$. By controlling
the time dependence of the parameter $\gamma(t)$, we aim to maximize
the probability for our system to be in the highest excited state $n$
of $H_{0}$ at a fixed time $T$.

For simplicity, we adopt the bang-bang control protocol. We divide the
total control time $T$ into $N$ periods with the same duration
$\delta t = T/N$. In the $i$-th period with
$(i-1)\delta t\le t\le i \delta t$ ($1\le i\le N$), the coupling is
either switched on or switched off, i.e., $\gamma(t)=a_{i}\gamma$ with
$a_{i}\in\{0,1\}$. Then a control strategy is specified by a series of
binary numbers $\{a_{1},a_{2},\cdots,a_{N}\}$. We aim to find out an
optimal strategy to maximize the fidelity
\begin{equation}
  \label{eq:3}
  \mathcal{F} \left(\rho(T), \left| n\right\rangle\left\langle n\right|\right) =
  \langle n|\rho(T)|n\rangle.
\end{equation}
It is worthy to point out that, since the size of the set of the
strategy space is $2^{N}$, it is impossible to get the optimal
strategy by exhaustively searching in the strategy space for a large
$N$.

Note that we focus on a regime where $\gamma$ is much smaller than the
energy gap $E_n-E_1$, which implies that the probability of arriving
at the state $|n\rangle$ at any time is very small with the coupling
always on. However, the optimal strategy to improve the probability of arriving at the highest energy eigenstate of $H_{0}$ with
switching on/off the coupling $V$ can be understood as follows. First, we switch on the coupling $V$ for a short period
from a lower energy eigenstate to a higher energy eigenstate, then we switch off the coupling $V$ to avoid the effect of $|i\rangle\langle i+1|$. Further, when the coupling $V$ is switched off, the free Hamiltonian $H_{0}$ changes the state of the system while keeping the energy invariant. Thus the energy of the system can be increased by suitable arranges of switching on/off the coupling. 

In fact, we
will study the cases where the dimension of the Hilbert space is $4$,
$6$, $8$ and $10$ while we do not increase the number of the control
parameters, which brings a great challenge to get an optimal strategy
to arrive at the highest eigenenergy state by a sequence of jumps
$|1\rangle \leftrightarrow |2\rangle \leftrightarrow \cdots
\leftrightarrow |n\rangle$.

\section{Reinforcement learning: Actor Critic model \label{AC}}

To find out the optimal strategy in our multi-level quantum
control problem, we will adopt a modern reinforcement learning method
called the actor-critic model. In this section, we will give a short
review of the actor critic reinforcement learning model.

In the traditional reinforcement learning, there are two different
types of methods to implement artificial intelligence. One is the
value-based methods (such as the Q-learning~\cite{Watkins1989}), where
the agent learns the value function that maps each state-action pair
to a value. According to the value function, the agent will take the
action with the largest return value for each state. It works well
when the set of actions is finite. The other is the policy-based
methods (such as policy gradients~\cite{sutton}), where we directly
optimize the policy without using a value function. It is efficient
when the action space is continuous or stochastic.

\begin{figure}[htbp]
  \includegraphics[width=0.98\columnwidth{},keepaspectratio]{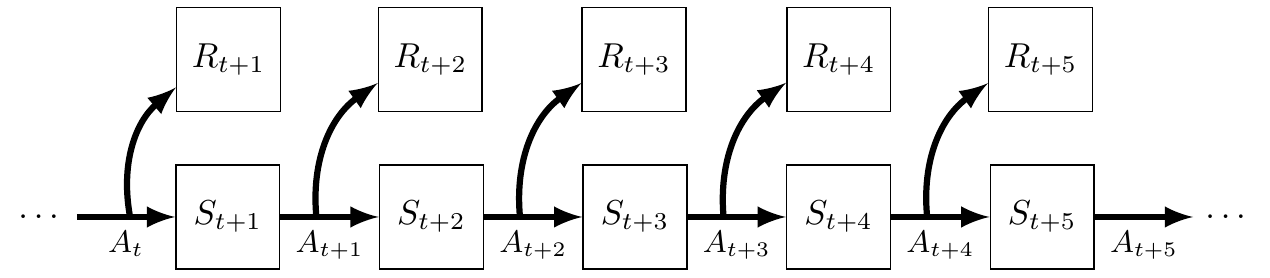}
  \caption{\label{fig:MDP}A schematic diagram of Markov decision
    process.}
\end{figure}

The reinforcement learning process is a finite Markov decision
process~\cite{sutton}. As shown in Fig.~\ref{fig:MDP}, a state $S_t$
at time $t$ is transmitted into a new state $S_{t+1}$ together with
giving a scalar reward $R_{t+1}$ at time $t+1$ by the action $A_t$
with the transmission probability $p(S_{t+1},R_{t+1}|S_t, A_t)$.

For a finite Markov decision process, the sets of the states, the
actions and the rewards are finite. In the value based methods, the
goal is to maximize the total discounted return at time $t$
\begin{equation}
  G_{t} = \sum_{k=0}^{\infty} \Gamma^{k} R_{t+k+1},
\end{equation}
where $\Gamma$ is the discount rate and $0\le\Gamma\le1$. The policy
$\pi$ is defined by the conditional probability $\pi(a|s)$ of
selecting an action $a$ for each state $s$. To estimate how good a
policy $\pi$ is, two value functions are introduced:
\begin{align}
  q_{\pi}(s, a) & \doteq \mathbb{E}_{\pi}\left[G_{t} | S_{t}=s, A_{t}=a\right],
  \\
  v_{\pi}(s) & \doteq \mathbb{E}_{\pi}\left[G_{t} | S_{t}=s\right],
\end{align}
where $q_{\pi}(s,a)$ is called the state-action value function,
$v_{\pi}(s)$ is called the state value function; $E_{\pi}$ denotes the
probability expectation for all the actions in the process taken
following the policy $\pi$. Note that we have the following relations:
\begin{align}
  q_{\pi}(s,a) & = \sum_{R} R p(R|s,a) + \Gamma \sum_{s^{\prime}} 
  v_{\pi}(s^{\prime})
               p(s^{\prime}|s,a), \\
  v_{\pi}(s) &  = \sum_{R,a}  R p(R|s,a) \pi(a|s) + \Gamma 
  \sum_{s^{\prime},a} v_{\pi}(s^{\prime})
               p(s^{\prime}|s,a) \pi(a|s). 
\end{align}
In addition, the advantage function is defined as
$A_{\pi}(s,a)=q_{\pi}(s,a)-v_{\pi}(s)$, which measures the advantage of an
action $a$ with respect to the state $s$ under the policy $\pi$.

In the policy gradient scheme, the objective is to maximize the
cumulant reward under a parameterized policy $\pi_{\theta}$:
\begin{equation}
  J\left(\pi_{\theta}\right) = 
  \mathbb{E}_{\pi_{\theta}}\left[\sum_{t=0}^{\infty} \Gamma^{t} 
  R\left(s_{t}\right)\right].
  \label{eq:pg}
\end{equation}
The model-free policy gradient of the cumulant reward is given by~\cite{SchulmanMLJA15}
\begin{equation}
  \nabla_{\theta} J\left(\pi_{\theta}\right)  \propto  \sum_{s} 
  \mu(s) \sum_{a} A_{\pi_{\theta}} \left(s,
                              a\right) \nabla_{\theta}  \pi_\theta\left(a
                              |s\right),\label{eq:4} 
\end{equation}
where $\mu(s)$ is the probability of appearing state $s$ in the Markov
process under the policy $\pi$. The above gradient can be estimated by
the score function estimator~\cite{estimator}.

\begin{figure}[htbp]
	\includegraphics[width=0.98\columnwidth{},keepaspectratio]{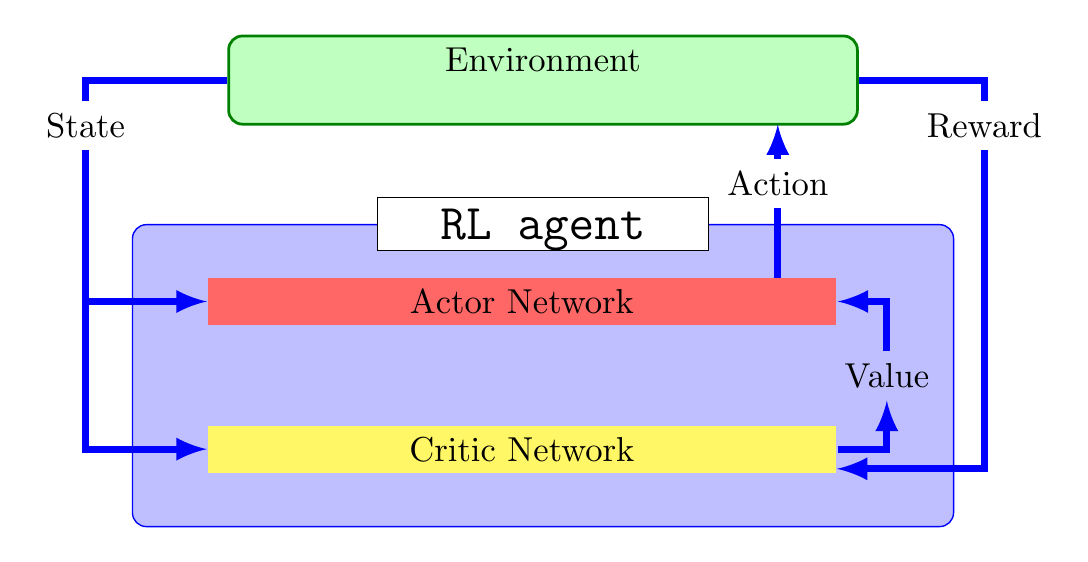}
	\caption{\label{fig:ac}A schematic diagram of actor-critic
          model: at each time step of training, the Actor network of
          the agent proposes a control action of $A_t$, the
          environment takes the proposed action and evaluates quantum
          state for time duration $\delta t$ to obtain the reward,
          both of which are fed into the RL agent. The Critic network
          of the agent receive the reward and estimate the action's
          value based on the state.}
            \end{figure}

In this paper, we use a hybrid type of reinforcement learning method,
called the actor-critic, whose protocol is shown in Fig.~\ref{fig:ac}.
The agent has two parts: a critic that measures how good the action
taken is and an actor that controls how our agent behaves. The actor
builds a network to evaluate the policy $\pi_{\theta}$, and takes an
action for the current state of the environment following the policy
$\pi_{\theta}$. The critic builds a network to evaluate the state
value function $v_{\phi}(s)$, which is used to approximate
$A_{\pi_{\theta}}(s,a)$ in Eq.~(\ref{eq:4}). The critic improves the
value network according to the reward from the environment, and the
actor improves the policy network according to a modified version of
Eq.~(\ref{eq:4}):
\begin{equation}
  \label{eq:5}
    \nabla_{\theta} J\left(\pi_{\theta}\right)  \propto  \sum_{s} \mu(s) \sum_{a} A_{\phi} \left(s,
                              a\right) \nabla_{\theta}  \pi_\theta\left(a
                              |s\right).
                          \end{equation}

In the actor-critic model, we get the advantage by building a network,
which is more efficient than by directly simulating following the
policy $\pi_{\theta}$. Besides, it improves the convergence
significantly to use the advantage function to replace the
state-action value function in evaluating the policy
gradient~\cite{a3c}.

In this work, we use the distributed proximal policy optimization
algorithm (DPPO)~\cite{DPPO2} to learn an optimal policy under the
policy gradient framework. The loss function of DPPO reads
\begin{widetext}
  \begin{eqnarray}
    L(\theta,\phi) = \hat{\mathbb{E}}_{\pi_{\theta_{\text{old}}}}
    \left[ \min \left( r_{\theta_{\text{old}}}(a|s_{0}, \theta)
    A_{\phi}(s_{0},a),  \operatorname{clip}
    \left(r_{\theta_{\text{old}}}(a|s_{0}, \theta), 1-\epsilon, 1+\epsilon \right)
    A_{\phi}(s_{0},a)\right)\right], 
  \end{eqnarray}
\end{widetext}
where $\epsilon$ is a hyper-parameter ($\epsilon=0.2$ in this paper).
The expectation $ \hat{\mathbb{E}}_{\pi_{\theta_{\text{old}}}}$
indicates the empirical average over a finite batch of samples under
the policy $\pi_{\theta_{\text{old}}}$. The term
$r_{\theta_{\text{odd}}}(a|s,\theta)$ is defined as the ratio of
likelihoods
\begin{equation}
r_{\theta_{\text{old}}}(a|s, \theta) = \frac{\pi_{\theta}(a | s)}{\pi_{\theta_{old}}(a |s)}.
\end{equation}
The clip function for $c \le d$ is defined as
\begin{equation}
  \label{eq:6}
  \text{clip}(f(x), c, d) =
  \begin{cases}
    d, & \text{ if } f(x) > d, \\
    f(x), & \text{ if } c \le f(x) \le d, \\
    c, & \text{ if } f(x) < c.
  \end{cases}
\end{equation}
The clip function for $r_{\theta_{\text{old}}}(a|s,\theta)$ penalizes
large changes between nearest updates, which corresponds to the trust
region of the first order policy gradient. Based on the first-order
trust region search gradient descent, DPPO has a robust learning
process and can handle both discrete and continuous action spaces. A
detailed description of the DPPO can be found in the
Appendix~\ref{PPO}.

\section{Quantum state control with Actor-Critic learning}
\label{sec:quant-state-contr}

\subsection{Agent-environment interface}

\begin{figure}[htbp]
  \includegraphics[width=0.98\columnwidth{}]{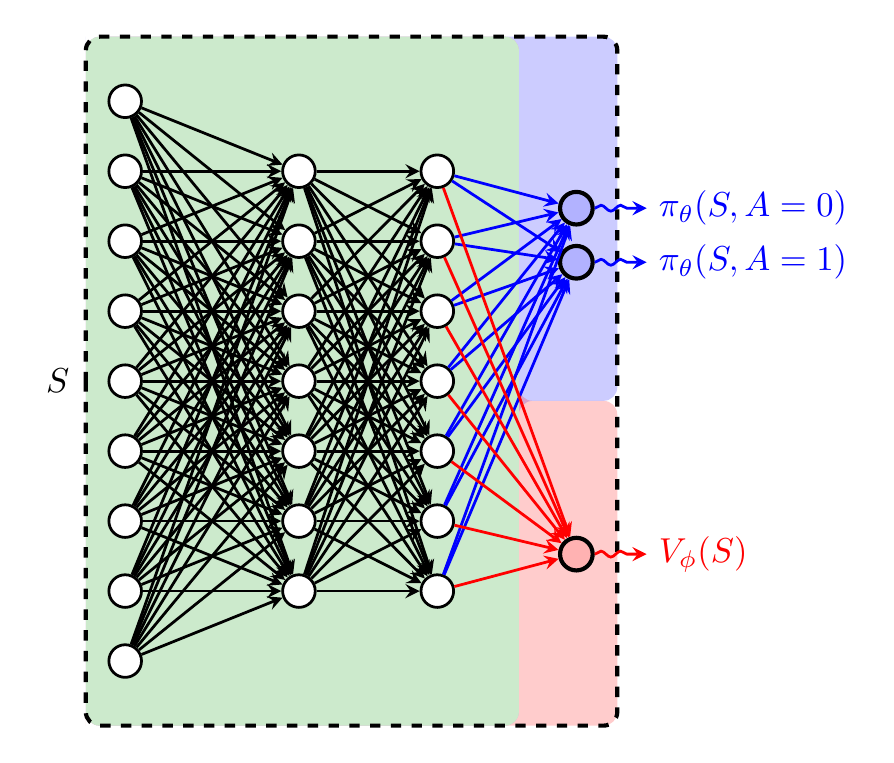}
  \caption{\label{fig:net} The architecture of the actor-critic
    neural network for the agent. The actor and the critic share the
    same architecture of hidden layers (green). The actor network has
    an action head (blue) to output the possible policy. The critic
    network has a value head (red) to output the value of the given
    state. }
\end{figure}

To implement the RL agent for our problem, we propose an interactive
interface between the RL agent and the physical environment
(Fig.~\ref{fig:ac}) adapted to OpenAI Gym~\cite{1606.01540}. We have
used Tensorflow~\cite{tensorflow} and Baselines~\cite{baselines} to
implement the learning algorithms with QuTip~\cite{qutip,qutip2}
simulating the dynamics of our control problem. The architecture of
deep neural network in our RL agent is shown in Fig.~\ref{fig:net}. In
our quantum control problem, the state at time $t$ in the
reinforcement learning is the state $\rho(t)$, which is expressed by its
components:
\begin{equation}
  \begin{aligned}
    s_{t}=\{& \Re(\rho_{11}(t)), \Im(\rho_{11}(t)),\\
    & \Re(\rho_{12}(t)), \Im(\rho_{12}(t)),\dots
    \\& \Re(\rho_{nn}(t)), \Im(\rho_{nn}(t))\},
  \end{aligned}
\end{equation}
where $\Re(\rho_{ij}(t))$ and $\Im(\rho_{ij}(t))$ are the real and the imaginary
part of the component $\rho_{ij}(t)$ respectively. Our action space is
formed by a switchable control field $a_{t} \in \{0,1\}$, which steers
our quantum state $\rho(t)$ to $\rho(t+\delta t)$ according to
Eq.~(\ref{eq:evo}). After evaluating the new state $\rho(t+\delta t)$ the
agent obtains the single step reward
\begin{equation}
  \label{eq:7}
  R_{t+1} = \mathcal{F}(\rho(t+\delta t),|n\rangle\langle n|) - \mathcal{F}(\rho(t),|n\rangle\langle n|), 
\end{equation}
where $\mathcal{F}$ is the fidelity defined by Eq. (\ref{eq:3}).

\subsection{Numerical results \label{result}}

We now apply the actor-critic RL approach to our quantum state control
problem with different settings, illustrating the flexibility and
efficiency of our RL agent. Here the different settings include
different numbers of energy level for our system, and different types
of environments affecting our system. We will give the numerical
results of the best fidelity $\mathcal{F}$ in our quantum state
control problem from the deep reinforcement learning. To show the
effectiveness of our deep RL method, we also calculate the fidelity
with the greedy method and gradient ascent pulse engineering (GRAPE) algorithm.
The greedy algorithm are used for finding successful policies by performing local searches.
The GRAPE method looks at the direct gradient of the fidelity function. 
In particular, to get better results, the GRAPE algorithm allows for the
coupling strength $\gamma(t)$ to take any value in the interval $[0, \gamma]$.
We then present our analysis of the
performance of our deep RL algorithm against the two algorithms.
Details of the greedy algorithm can be found in the appendixes.

\subsubsection{Quantum state control without environments}
\label{sec:pure}

In this subsection, we consider our quantum state control problem with
a quantum system with negligible environments. In other words, we
assume that all the coefficients $\Gamma_{k,n}=0$.

\begin{figure}[htbp]
  \includegraphics[width=0.98\columnwidth{},keepaspectratio]{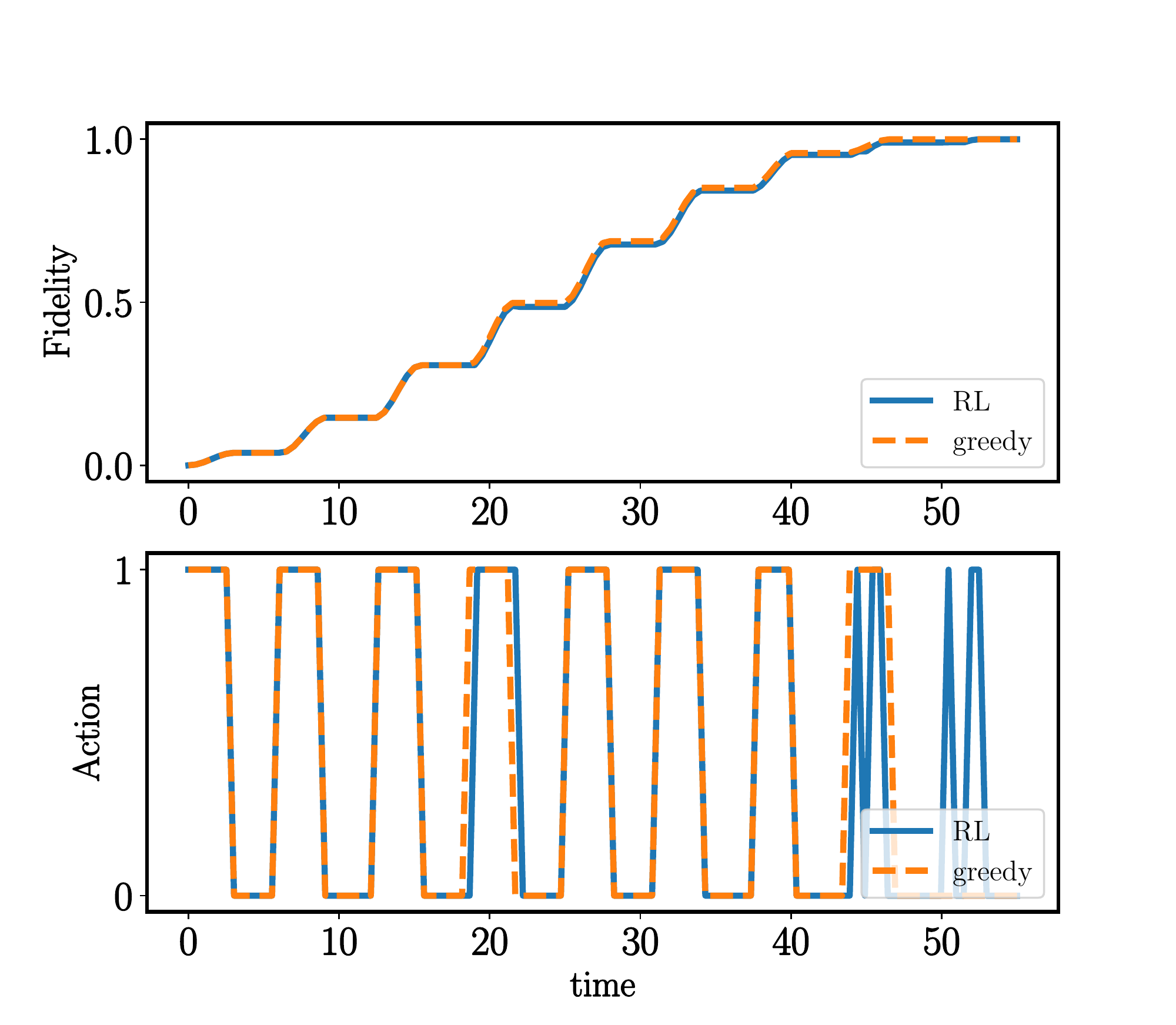}
  \caption{\label{fig:lev2} The best fidelities (up) and strategy
    (down) of preparing an excited state for the two-level control
    model with $\gamma=0.1$. The markers correspond to the algorithms RL
    (blue line) and greedy (orange line). The time step $N=110$. }
\end{figure}                                                                                                                          

In Fig.~\ref{fig:lev2} we show the results of the optimal fidelity and
the corresponding strategy on our quantum state control problem with
parameters $\{n=2,\, \gamma=0.1,\, T =55,\, N=110\}$ in Fig.~\ref{fig:lev2}.
With $1500$ episodes, our RL agent gets the optimal fidelity
$\mathcal{F}_{RL}(T)\approx 0.999998$, which is a little larger than the
fidelity $\mathcal{F}_{Greedy}(T)\approx 0.999815$ from the direct greedy
algorithm. While the difference of the fidelities between those two
methods is very small, the strategy in Fig.~\ref{fig:lev2} is
different for about $T>45$, which shows that our RL agent has learned
a globally optimized protocol in this task. Notice that for all
control tasks discussed in our manuscript, the time scale
$\delta t$ is always $0.5$. A detailed optimal strategy of greedy method
can be found in the Appendix~\ref{2lev}.

\begin{figure}[htbp]
	\centering
	\subfloat{
		\includegraphics[width=0.24\textwidth]{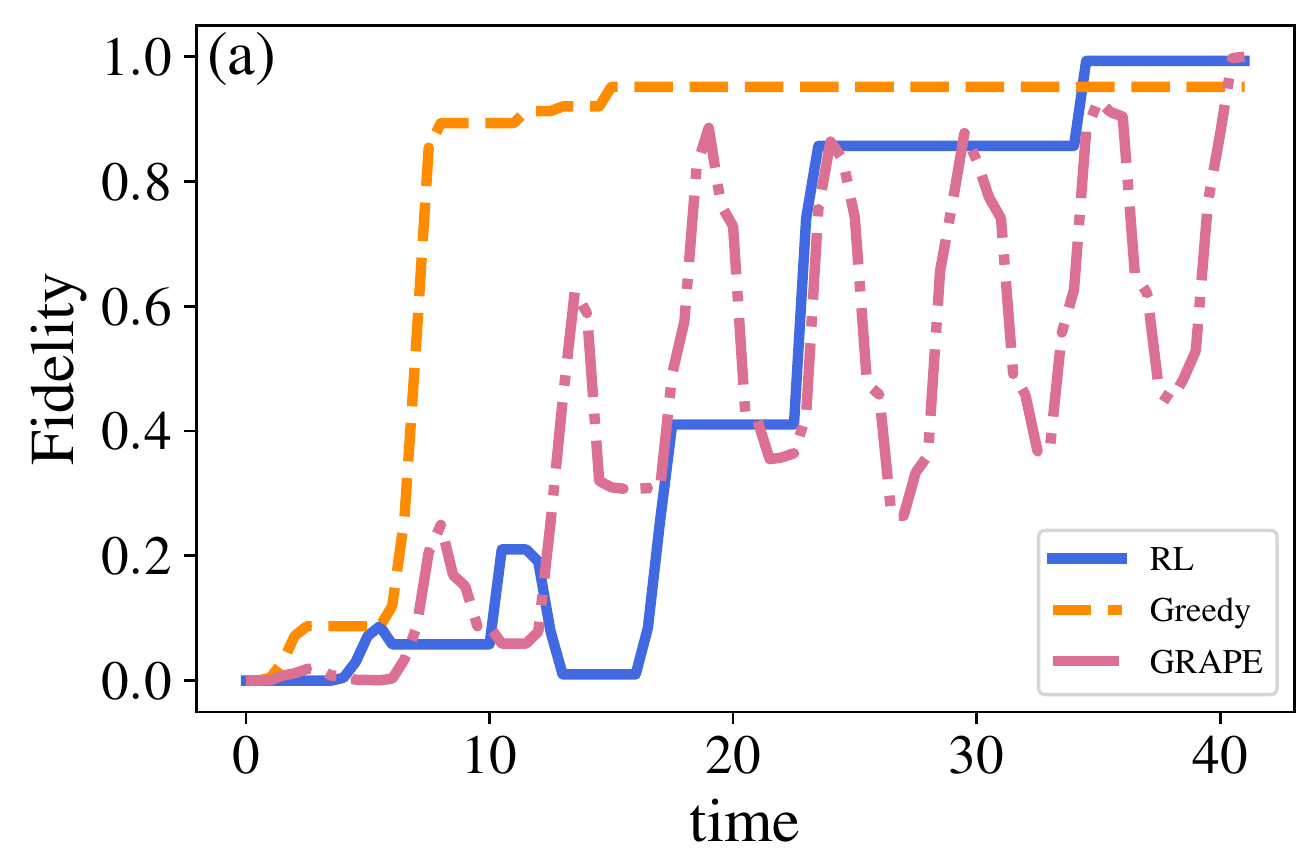}
	}
	\subfloat{
		\includegraphics[width=0.24\textwidth]{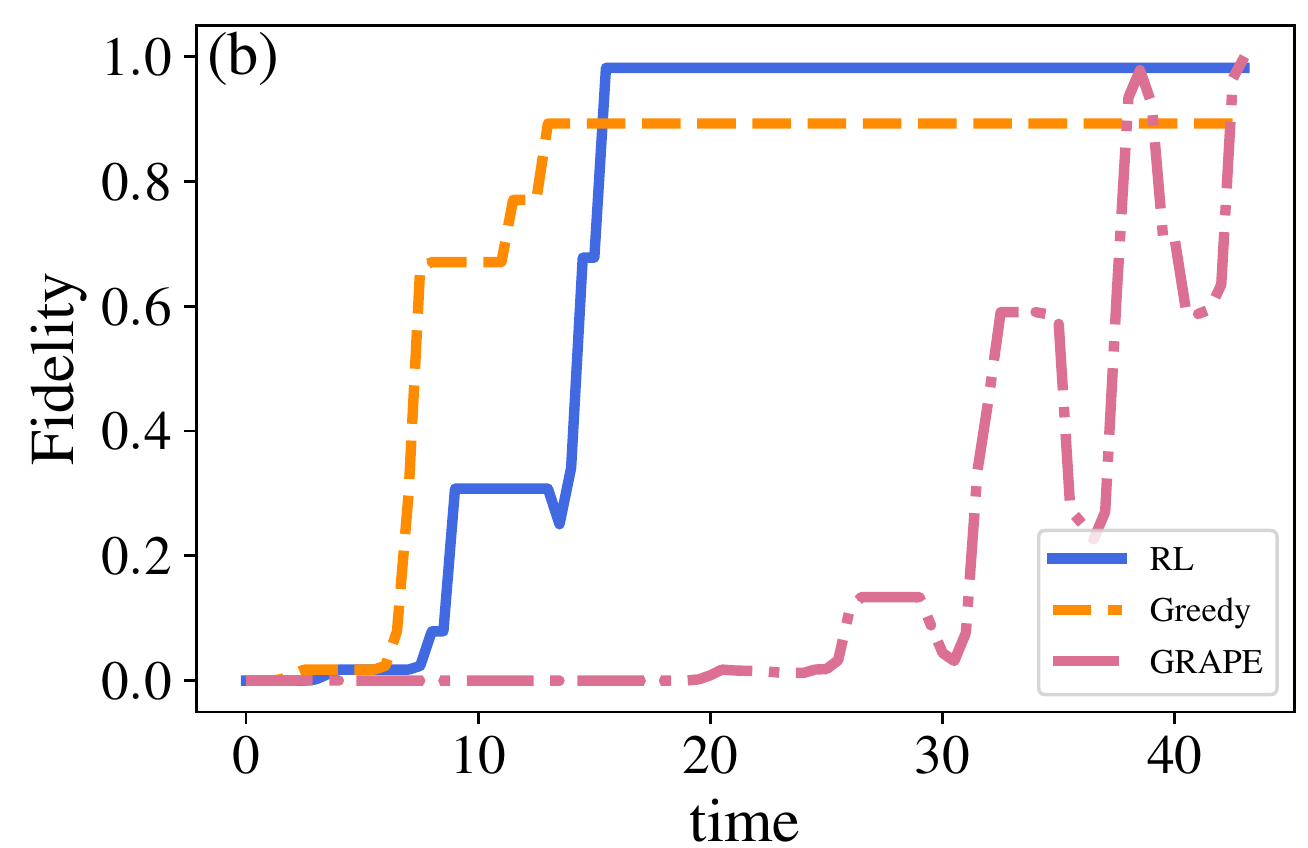}
	}
	\qquad
	\subfloat{
		\includegraphics[width=0.24\textwidth]{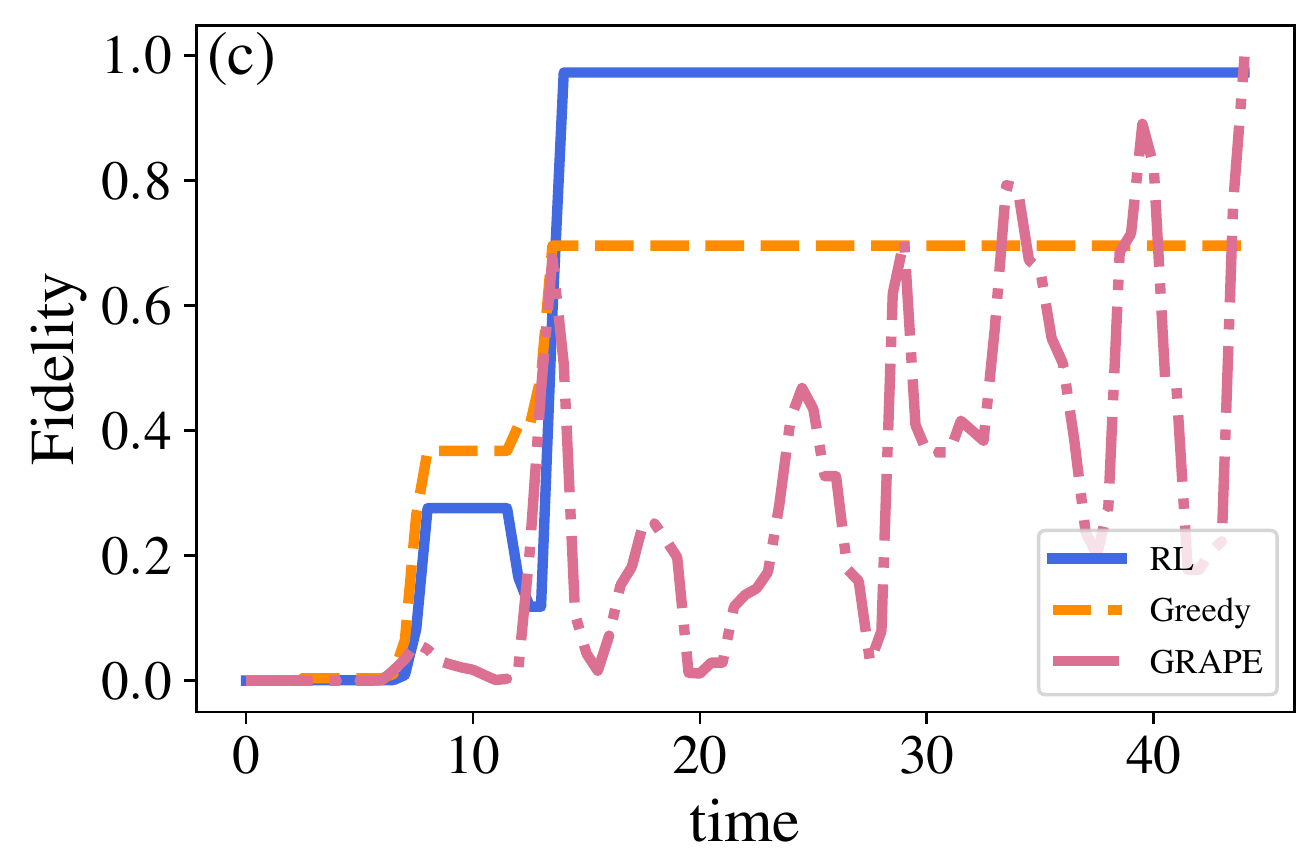}
	}
	\subfloat{
		\includegraphics[width=0.24\textwidth]{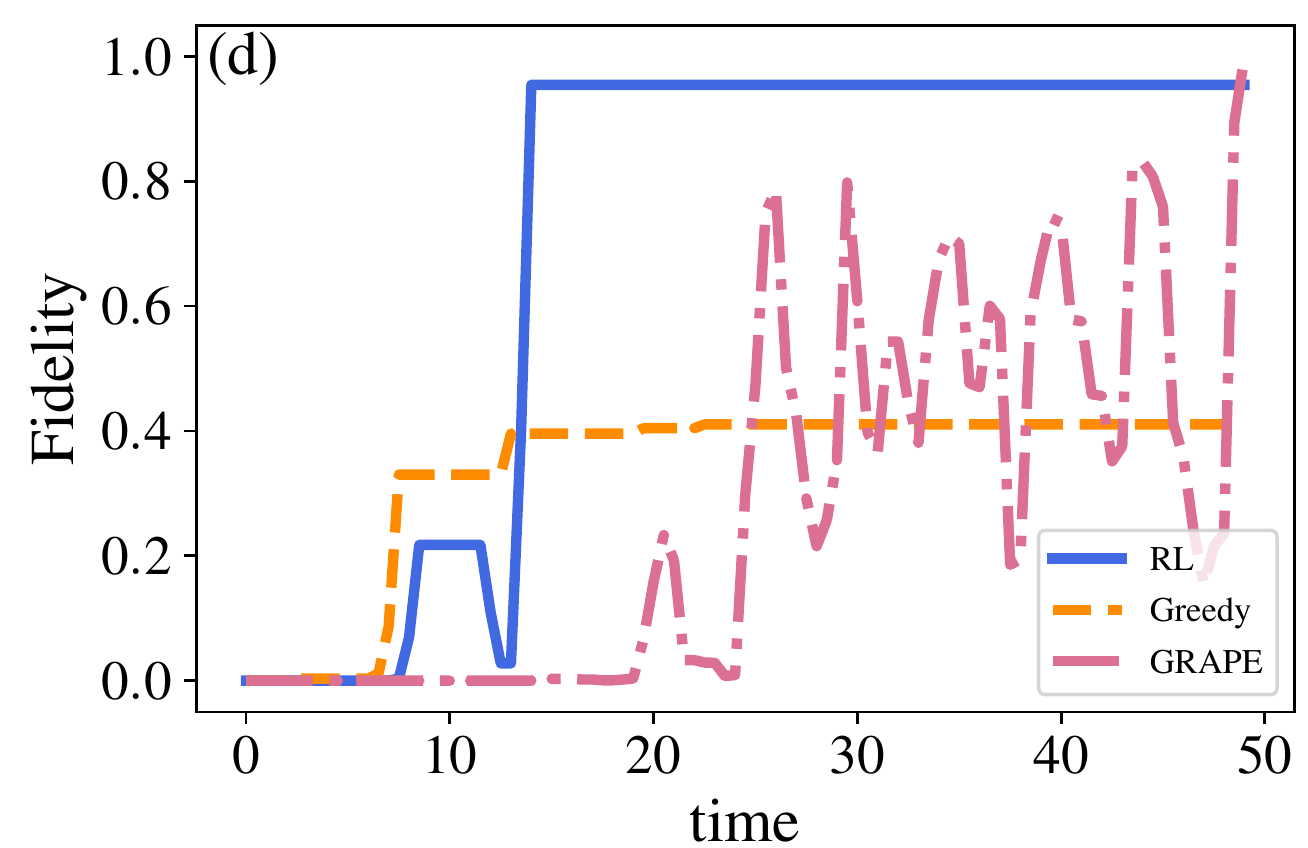}
	}
	\caption{Results from from the three algorithms
          for different level control model. The horizontal and
          vertical axes of each subfigure denote evolution time $t$
          and fidelity $\mathcal{F}$. (A),(B),(C),(D): The fidelities
          for three different methods with 4,6,8,10 level control model.
          The corresponding coupling
          strengths with the different models are $\gamma=0.8,1.1,1.4,1.9 $. The
          time steps with different control tasks are $N=82,86,88,98$. }
	\label{fig:levm}
\end{figure}

We further apply our RL agent to the quantum state control problem in
the multi-level Hilbert space. We give the optimal fidelities in the
cases with the dimension of Hilbert space equal to $4$, $6$, $8$, and
$10$ by the RL algorithm (red dashed line), the greedy algorithm
(blue solid line) and GRAPE (violet dot dashed line), which are shown in Fig.~\ref{fig:levm}(A)-(D). We
find that the greedy algorithm becomes less effective with the
increase of the dimension of Hilbert space, but the RL algorithm and GRAPE
performs well in all cases. For example, when the dimension of Hilbert
space varies from $4$ to $10$, the optimal fidelity from the greedy
algorithm varies from about $0.954$ to about $0.411$, but the fidelity
from the RL algorithm varies from about $0.993$ to $0.954$. While GRAPE has the best performance out of the three
methods, the algorithm requires the fidelity gradients at
all time. 

\subsubsection{Quantum state control with environments}

We now turn our attention to the behavior of our learning strategy
when applied to a non-ideal scenario in which typical realistic
conditions are considered. In particular, we discuss the results
produced by RL agent when the system is affected by dephasing and
energy decay.

\begin{figure}[htbp]
	\centering
	\subfloat{
		\includegraphics[width=0.24\textwidth]{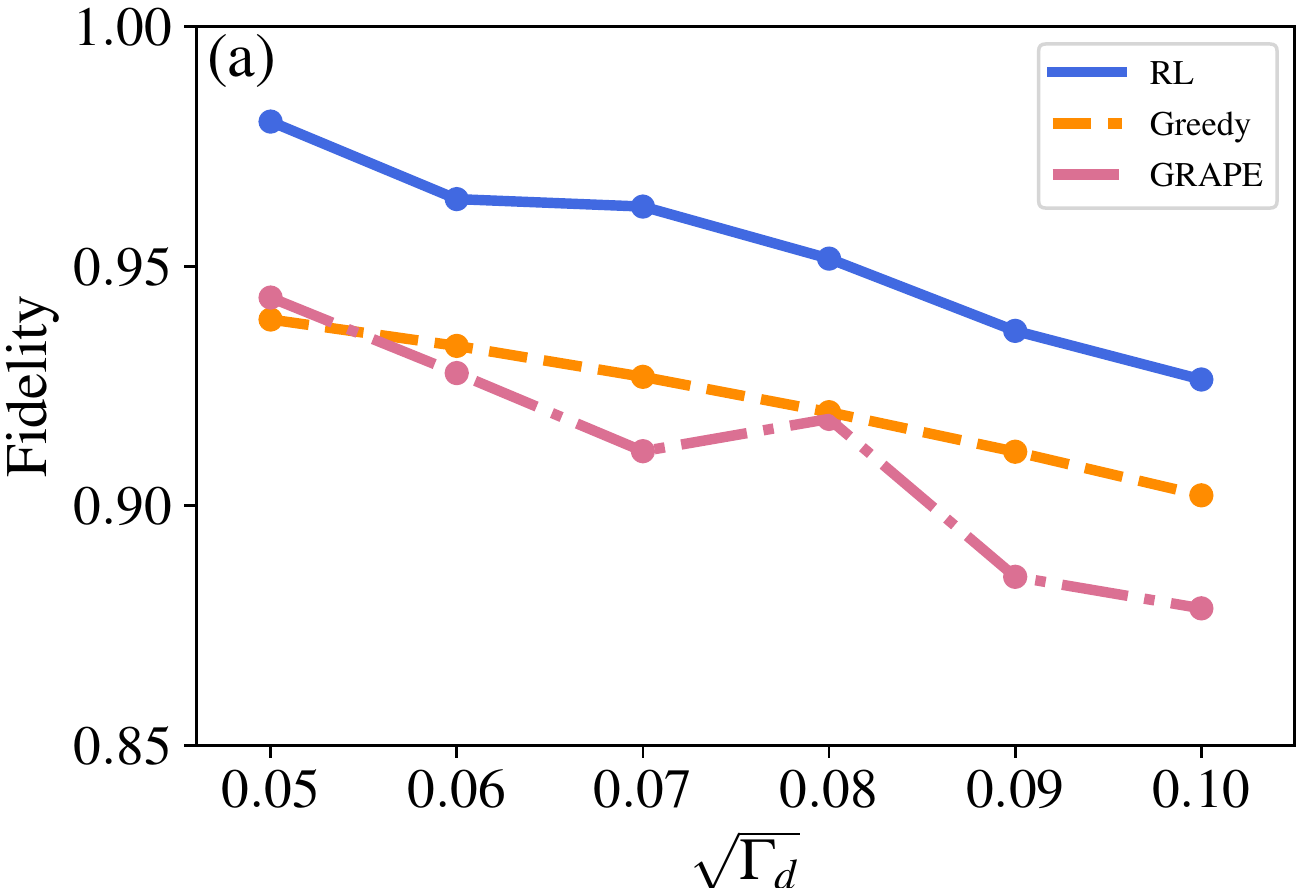}
	}
	\subfloat{
		\includegraphics[width=0.24\textwidth]{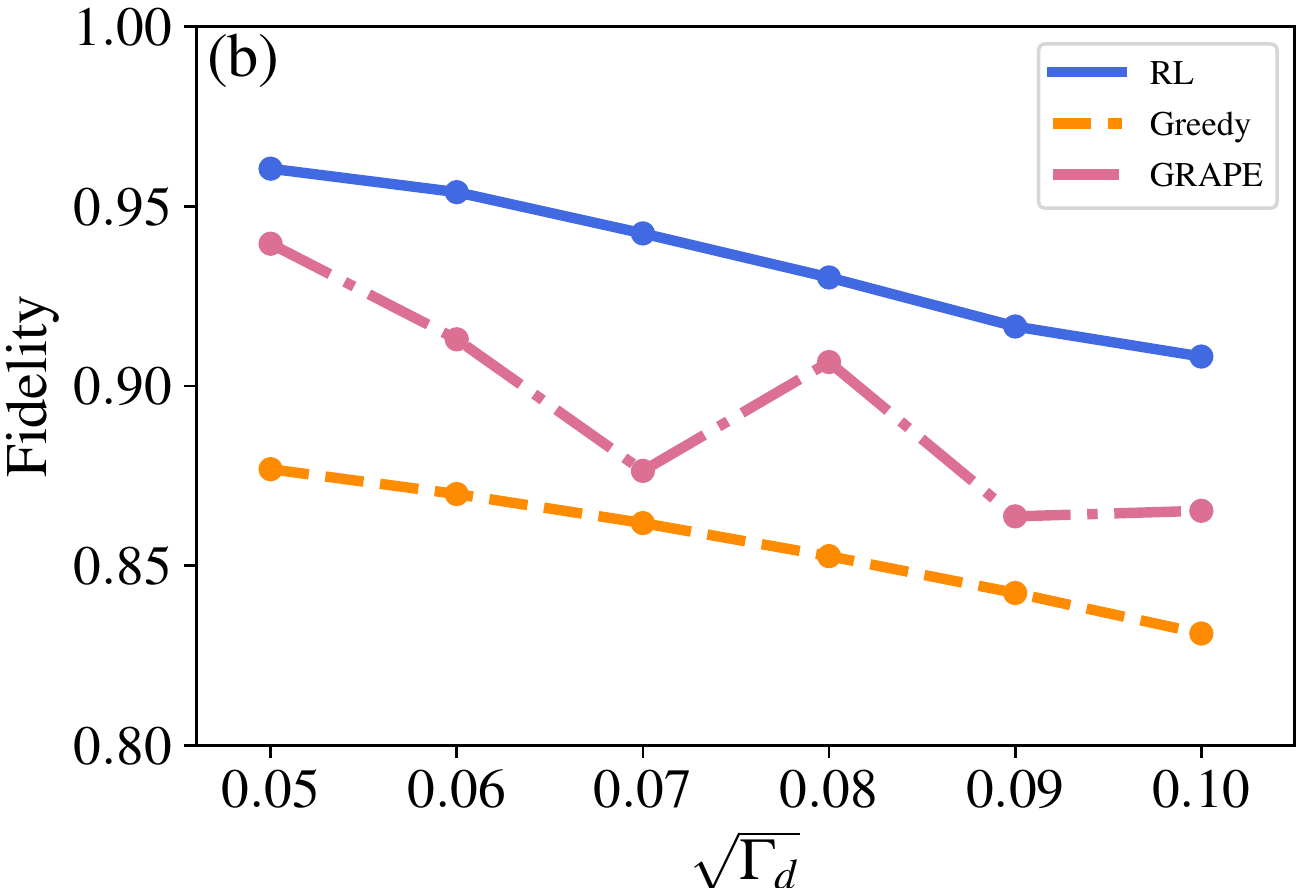}
	}
	\qquad
	\subfloat{
		\includegraphics[width=0.24\textwidth]{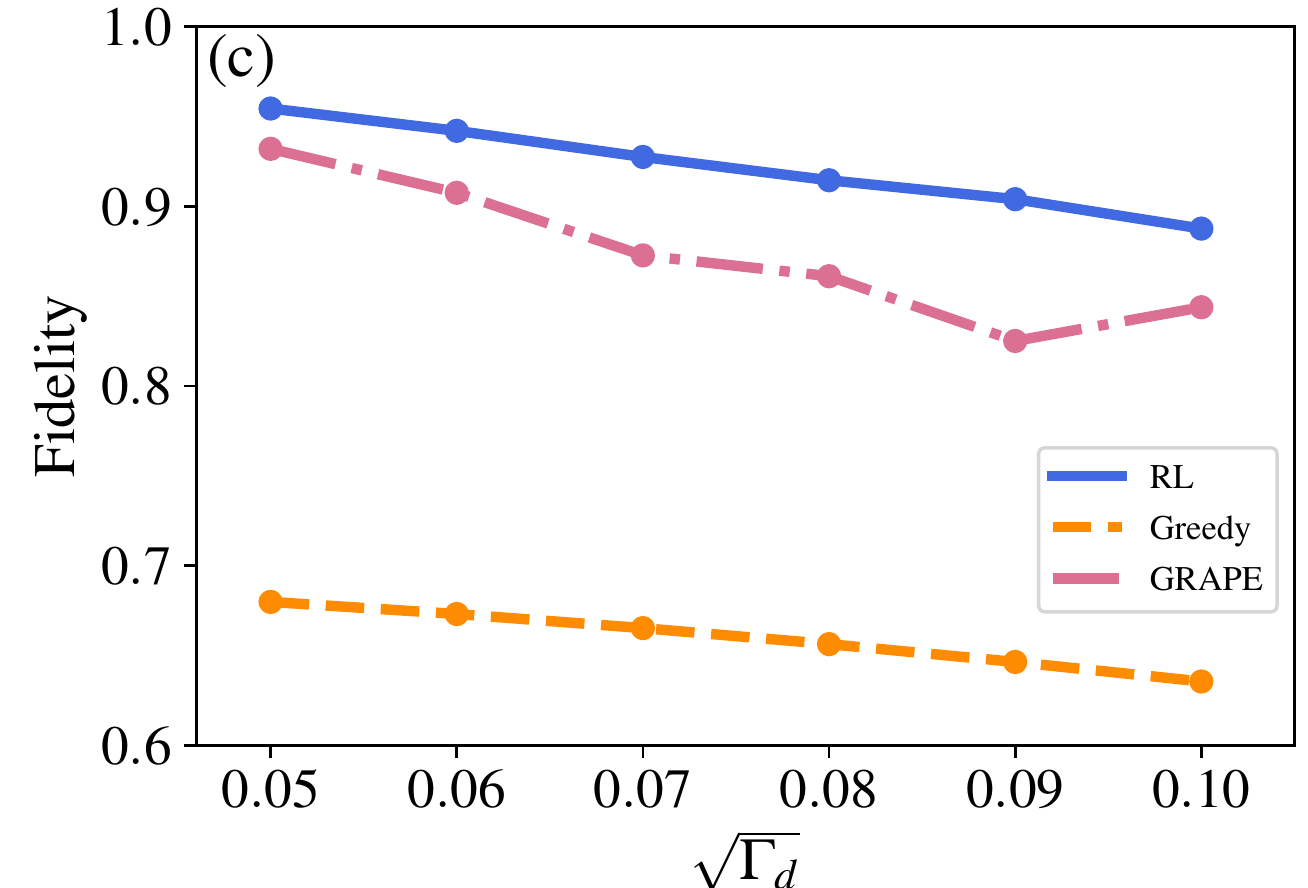}
	}
	\subfloat{
		\includegraphics[width=0.24\textwidth]{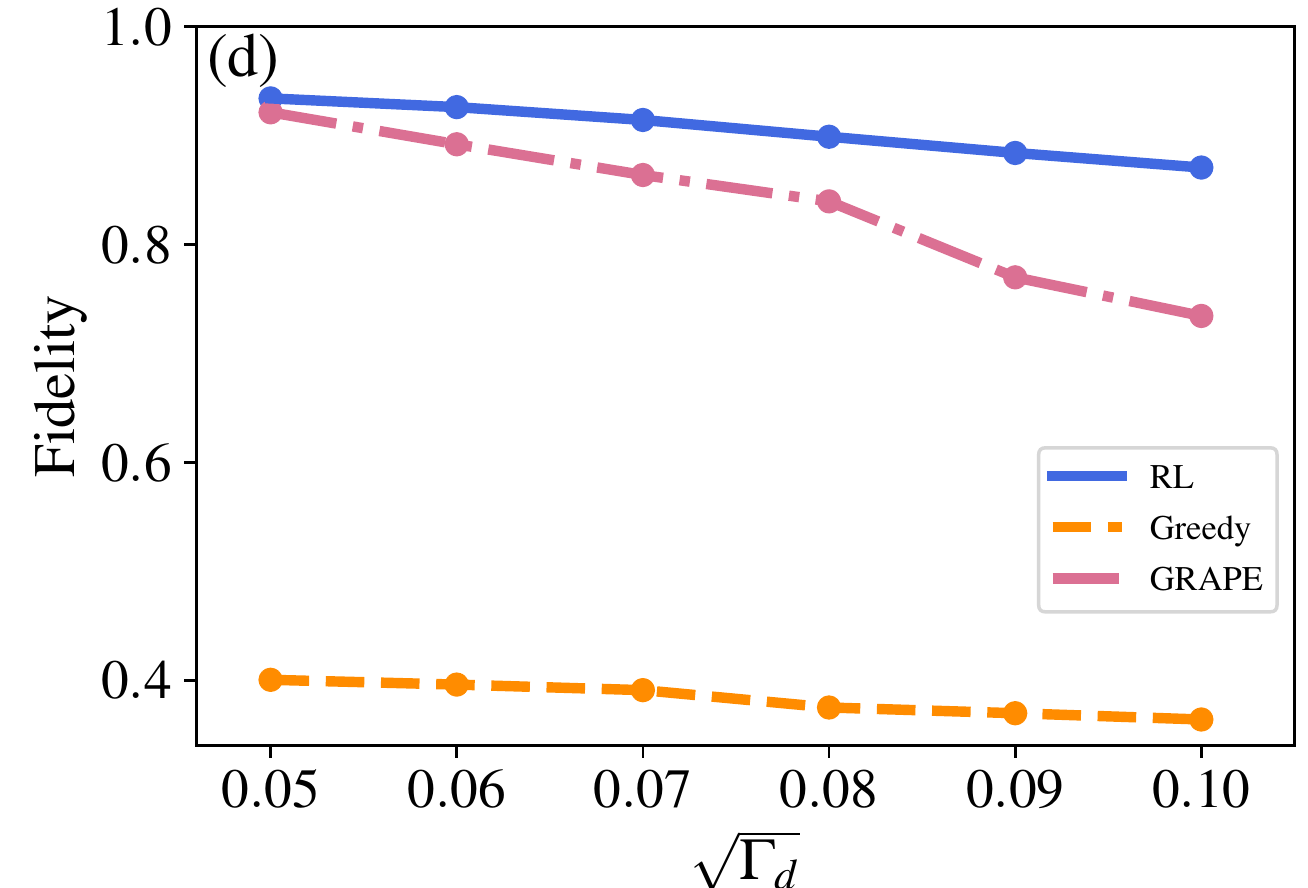}
	}
	\caption{Results from from the three algorithms
		for different level control model under dephasing 
		dynamics.  The horizontal and vertical axes of  each 
		subfigure denote dephasing rate $\sqrt{\Gamma_d}$ and 
		fidelity $\mathcal{F}$. 
		(A),(B),(C),(D): Best fidelity for three different methods of
		4,6,8,10 level control model. Note that the Hamiltonian is the
		same as Fig.~\ref{fig:levm} showed.}
	\label{fig:levd}
\end{figure}

In Fig.~\ref{fig:levd} we present our numerical results on the control
problem under dephasing dynamics. Fig.~\ref{fig:levd}(A)-(D) show the
results for dephasing rate
$\sqrt{\Gamma_{d}}=\{0.05,0.06,0.07,0.08,0.09,0.1\}$. In both cases, our
best results from the RL agent outperform the greedy algorithm and even GRAPE. Also, with the
energy level number getting higher, the differences of fidelities between
the three methods get larger.

\begin{figure}[htbp]
	\centering
	\subfloat{
		\includegraphics[width=0.24\textwidth]{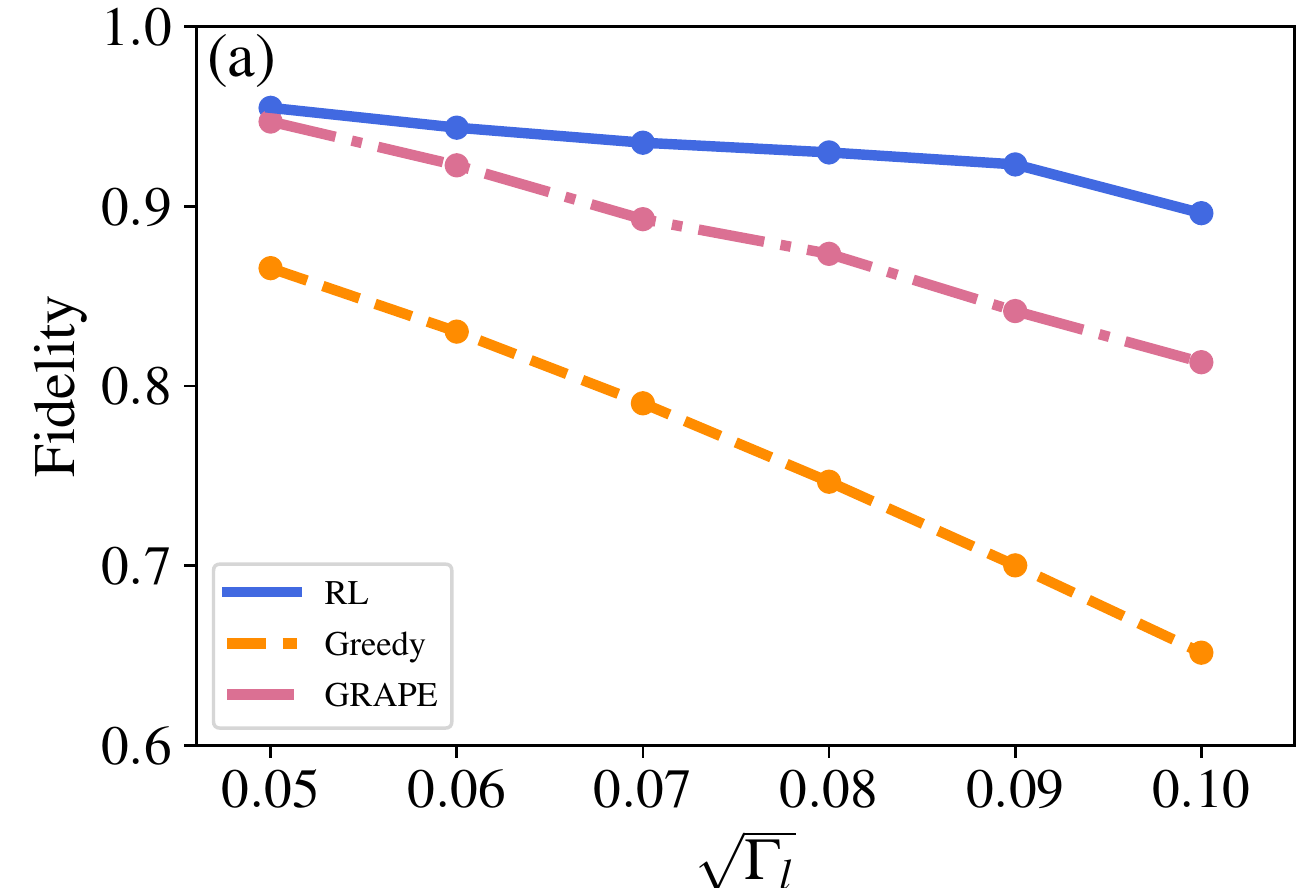}
		}
	\subfloat{
		\includegraphics[width=0.24\textwidth]{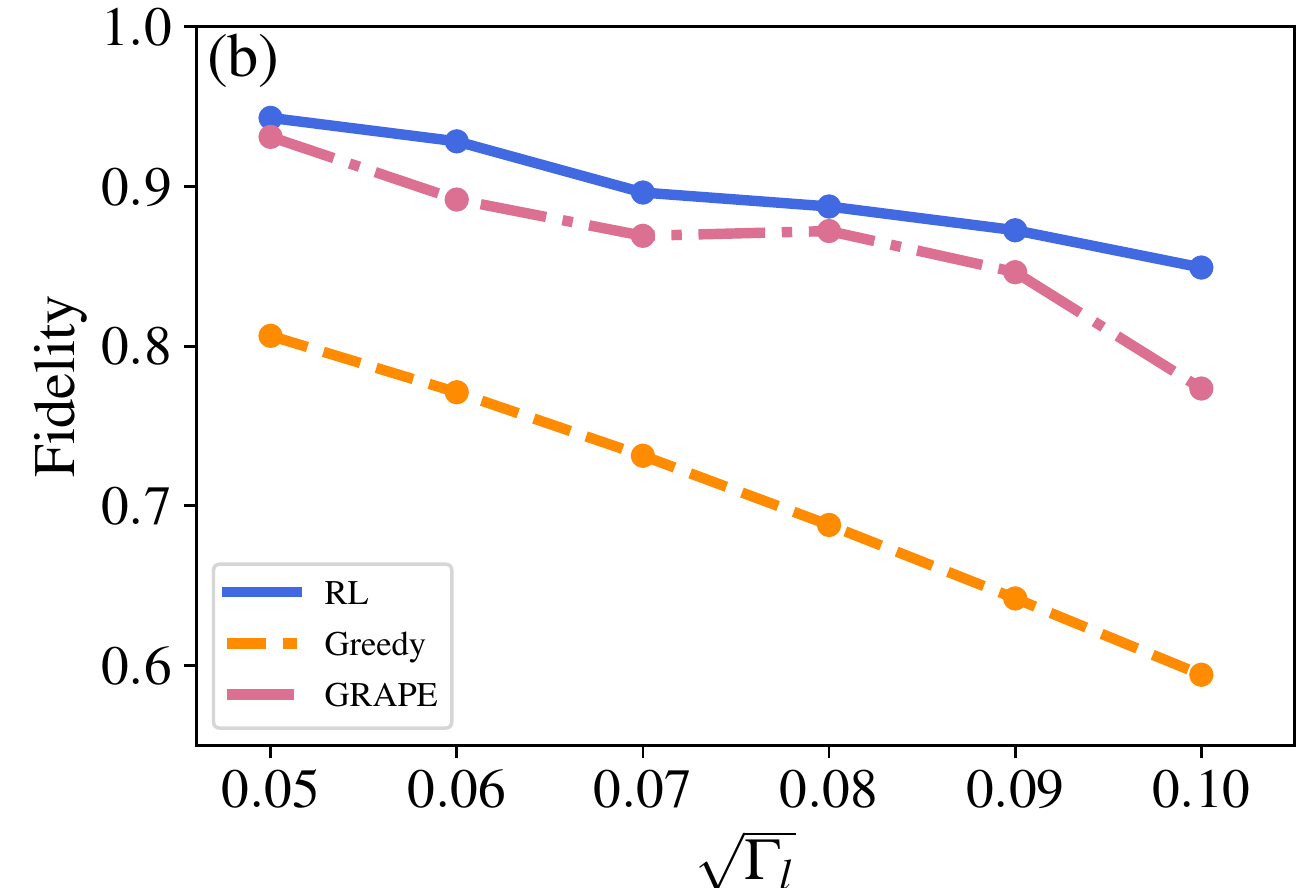}
		}
	\qquad
	\subfloat{
		\includegraphics[width=0.24\textwidth]{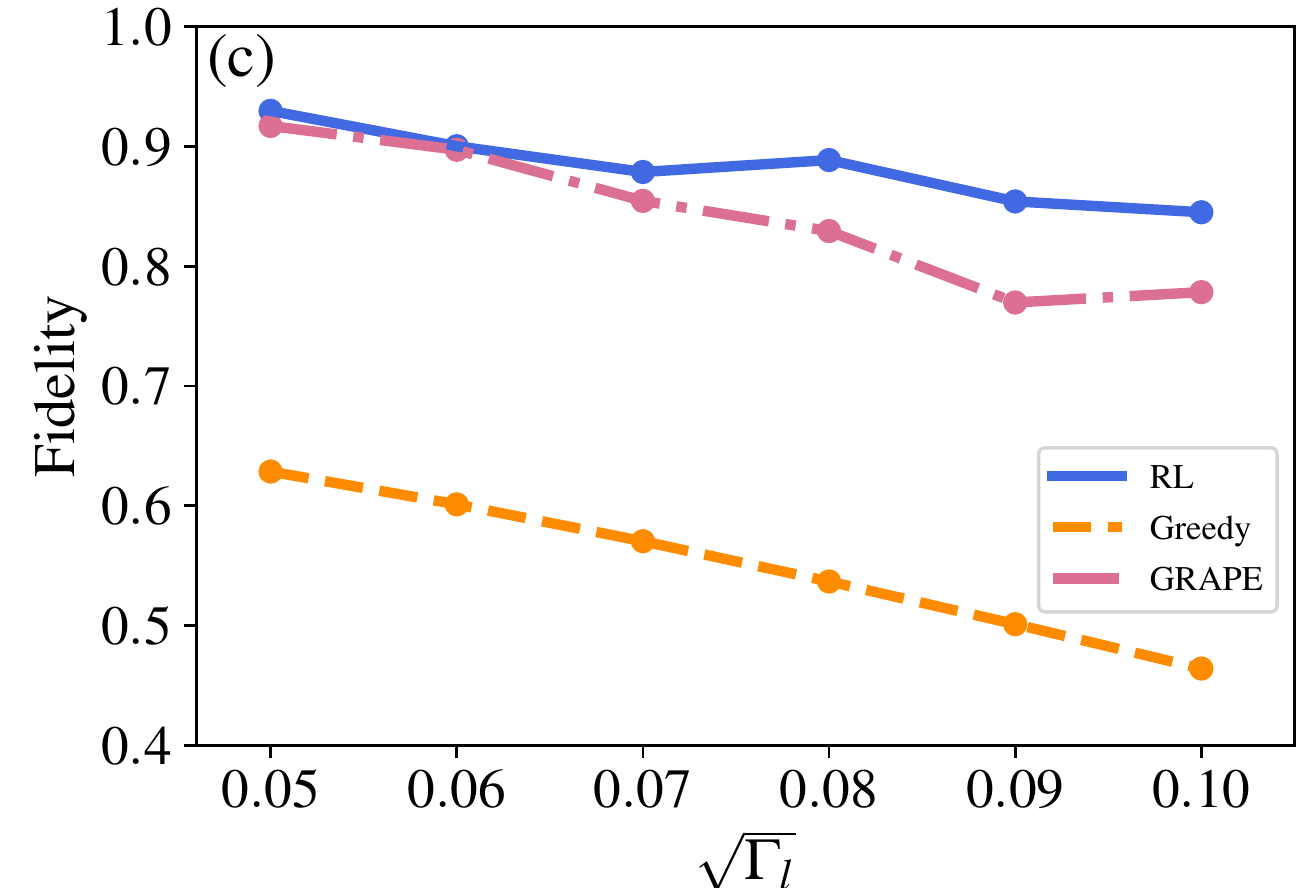}
		}
	\subfloat{
		\includegraphics[width=0.24\textwidth]{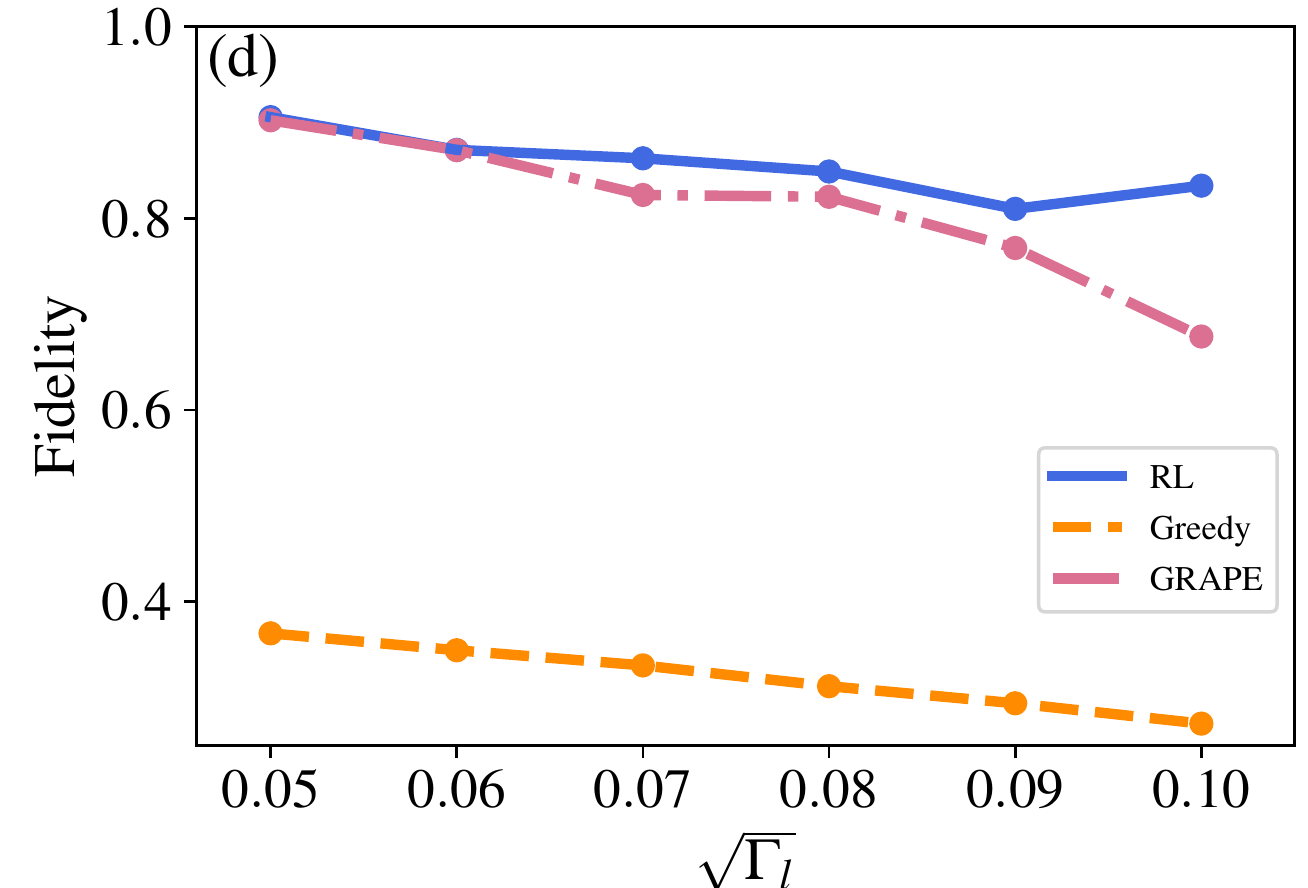}
	}
	
	\caption{Results from from the three algorithms
		for different level control model under energy decay dynamics.
		 The horizontal and vertical axes of  each subfigure denote 
		 energy decay rate $\sqrt{\Gamma_l}$ and fidelity 
		 $\mathcal{F}$. 
		(A),(B),(C),(D): Best fidelity for three different methods of
		4,6,8,10 level control model. Note that the Hamiltonian is the
		same as Fig.~\ref{fig:levm} showed.}
		\label{fig:levl}
\end{figure}

Fig.\ref{fig:levl} shows the superior performance of the RL agent
versus the greedy and GRAPE during the time evolution under the disturbance of
energy decay. Similar to the dephasing cases, the RL agent has successfully
conquered the control problem under energy decay dynamics.
While for the
greedy algorithm, it is impossible to get a convincing result with a
large energy decay rate in high dimensional control problems. 
In this scenario, we find that the RL agent successfully learns
to adapt to overcome disturbance of energy decay in multi-level control
problems.

\begin{figure}
	\centering
	\subfloat{
		\includegraphics[width=0.24\textwidth]{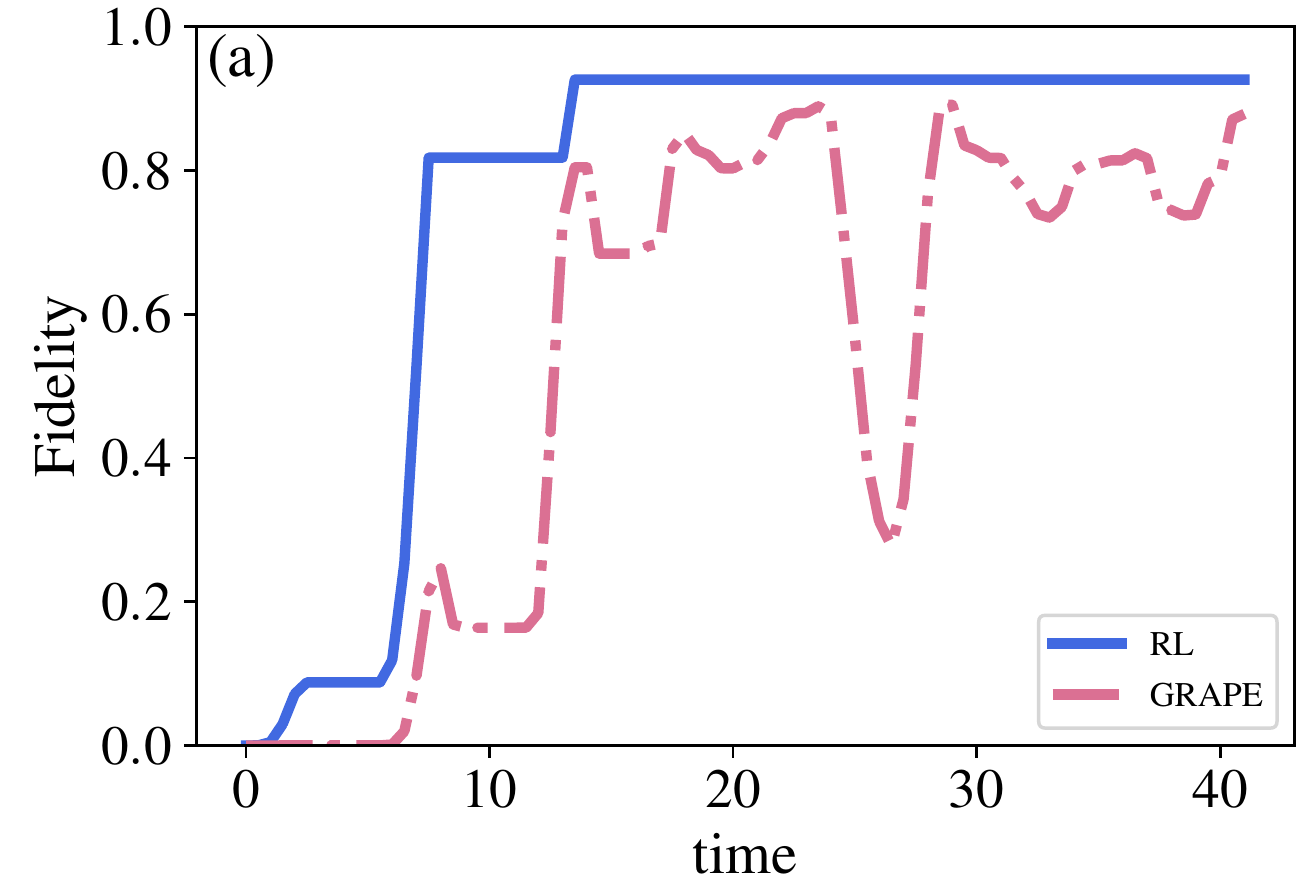}
	}
	\subfloat{
		\includegraphics[width=0.24\textwidth]{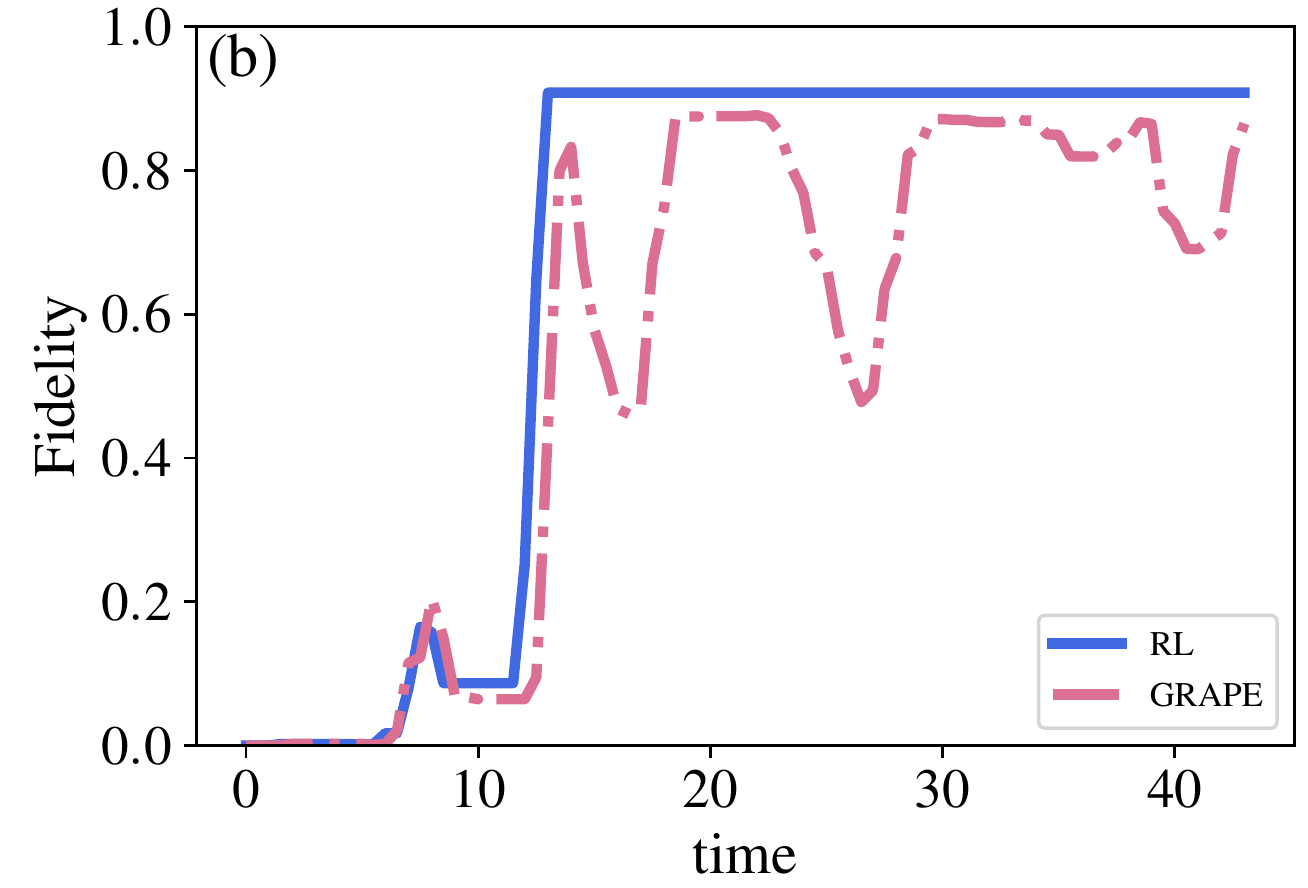}
	}
	\qquad
	\subfloat{
		\includegraphics[width=0.24\textwidth]{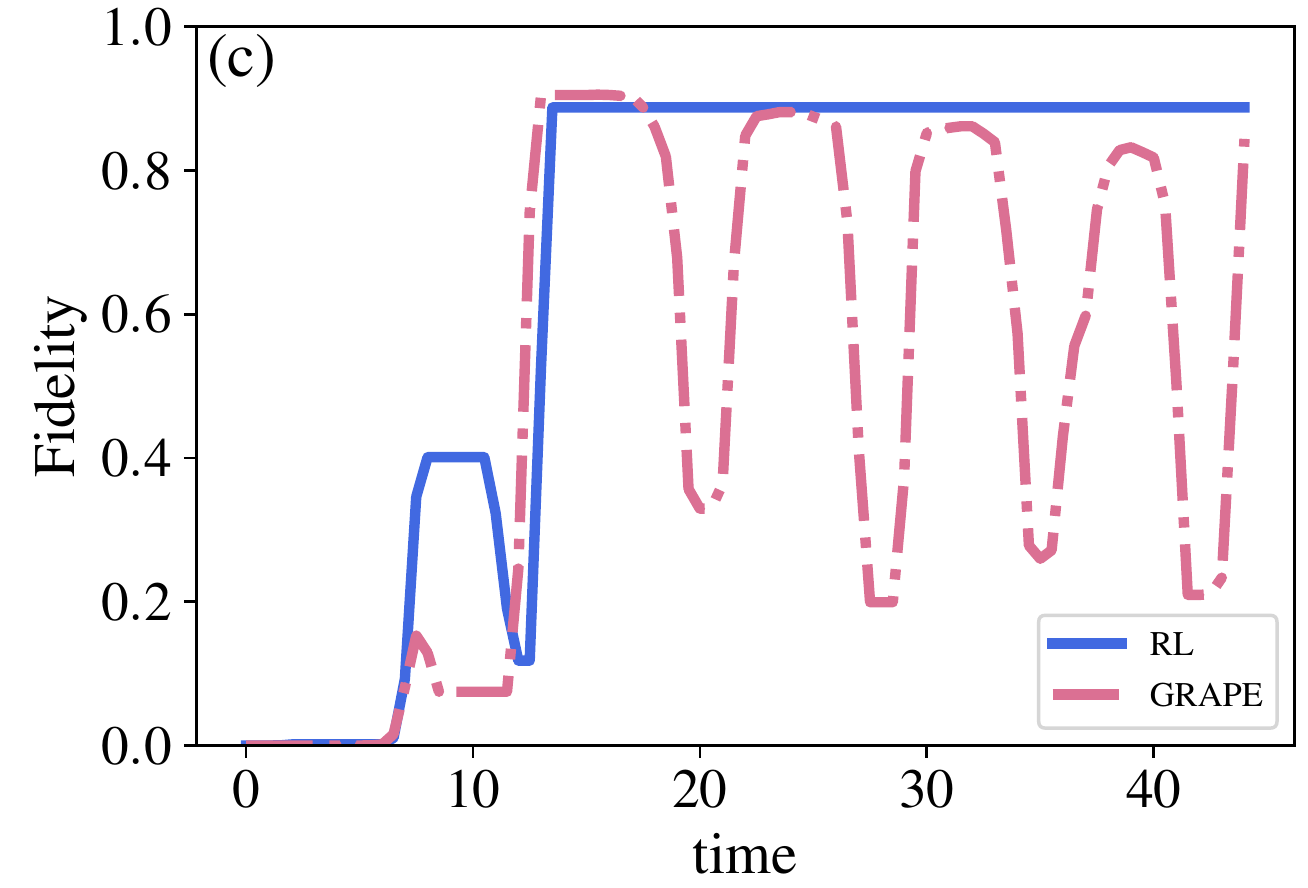}
	}
	\subfloat{
		\includegraphics[width=0.24\textwidth]{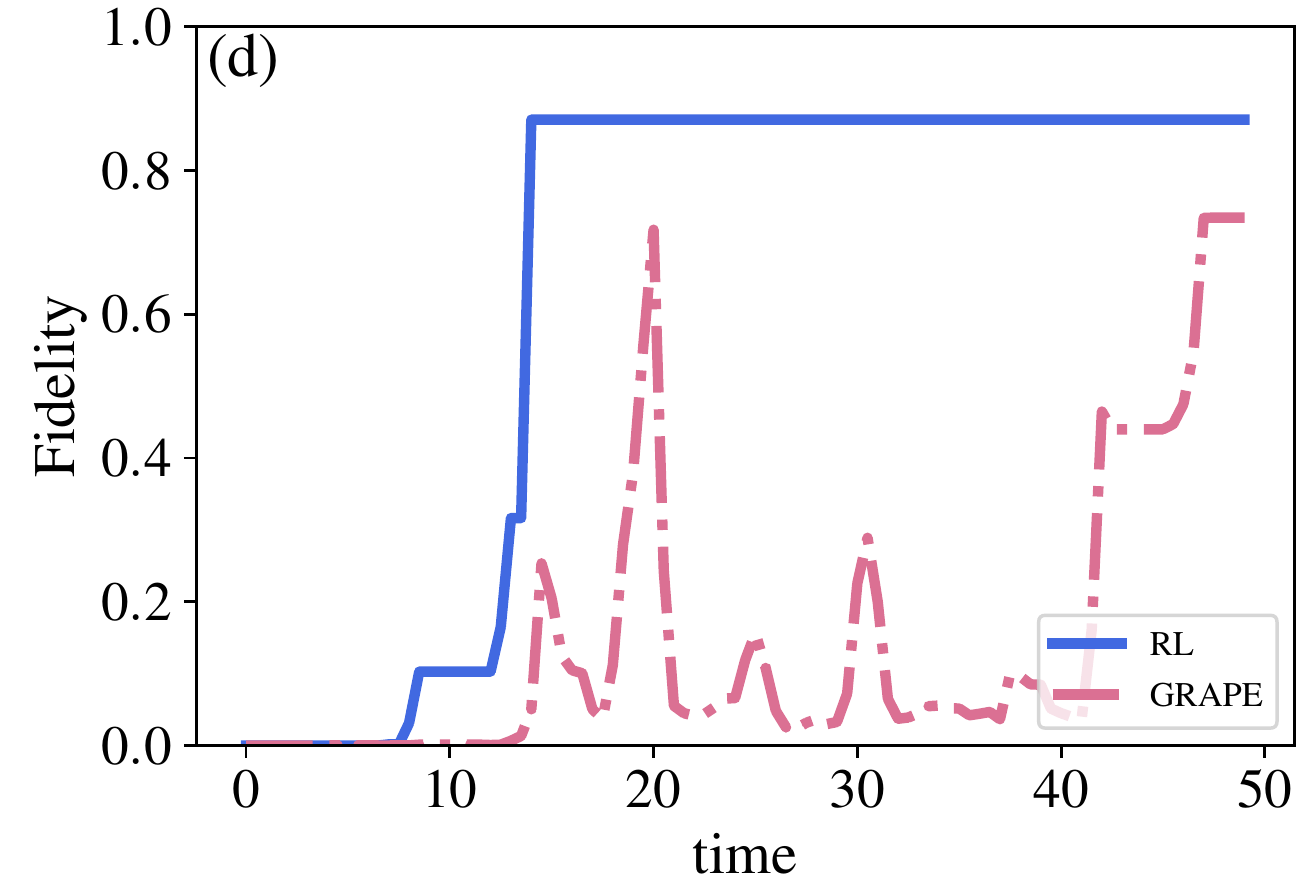}
	}
	\caption{Results from RL agent and GRAPE strategies 
		for different high level control model with 
		$\sqrt{\Gamma_d}=0.1$.The horizontal and vertical axes of 
		each subfigure denote evolution time $t$ and fidelity 
		$\mathcal{F}$.
		(A),(B),(C),(D): The evolution of fidelity with RL agent and GRAPE control 
		of
		4,6,8,10 level control model.}
	\label{fig:evod}
\end{figure}

\begin{figure}
	\centering
	\subfloat{
		\includegraphics[width=0.24\textwidth]{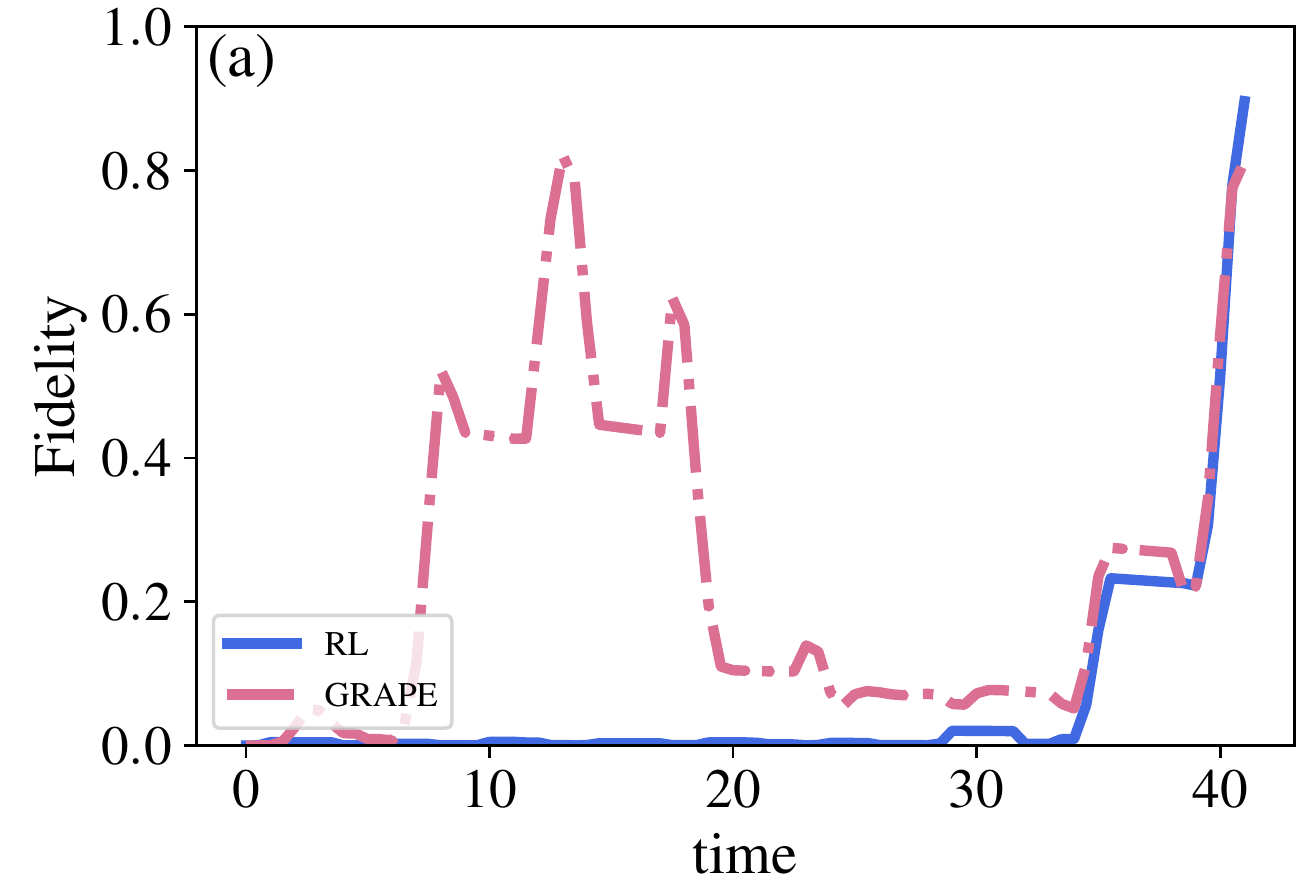}
		}
	\subfloat{
		\includegraphics[width=0.24\textwidth]{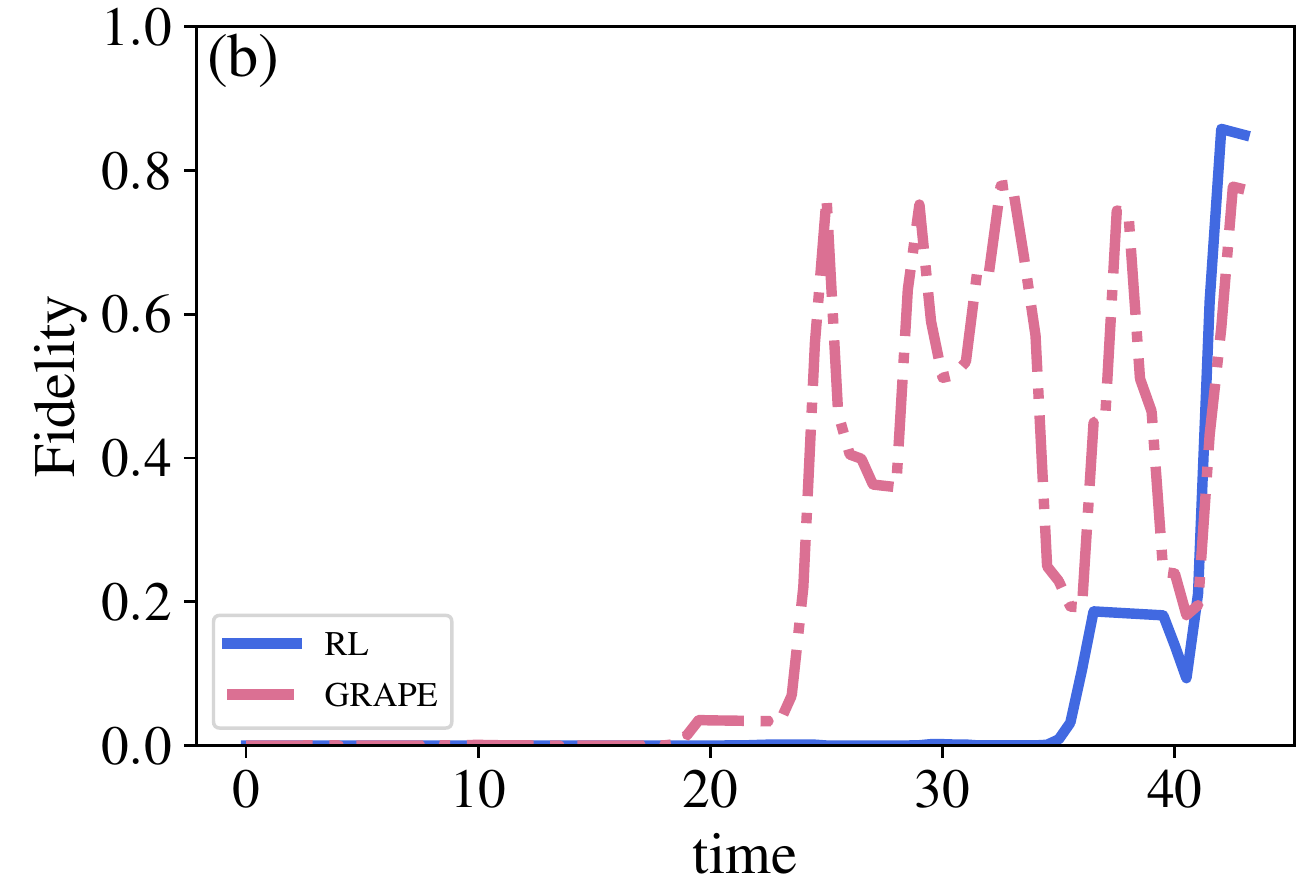}
		}
	\qquad
	\subfloat{
		\includegraphics[width=0.24\textwidth]{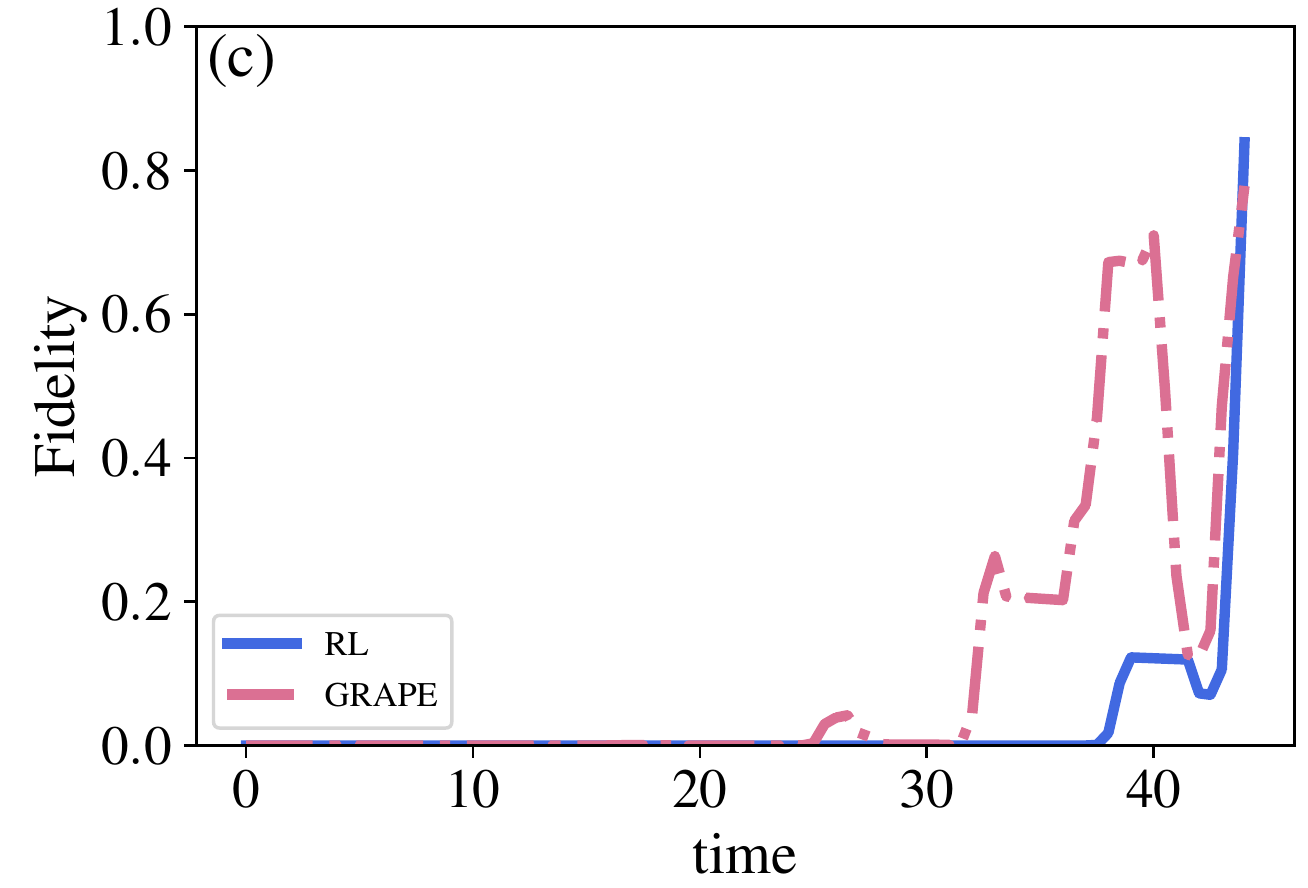}
	}
	\subfloat{
		\includegraphics[width=0.24\textwidth]{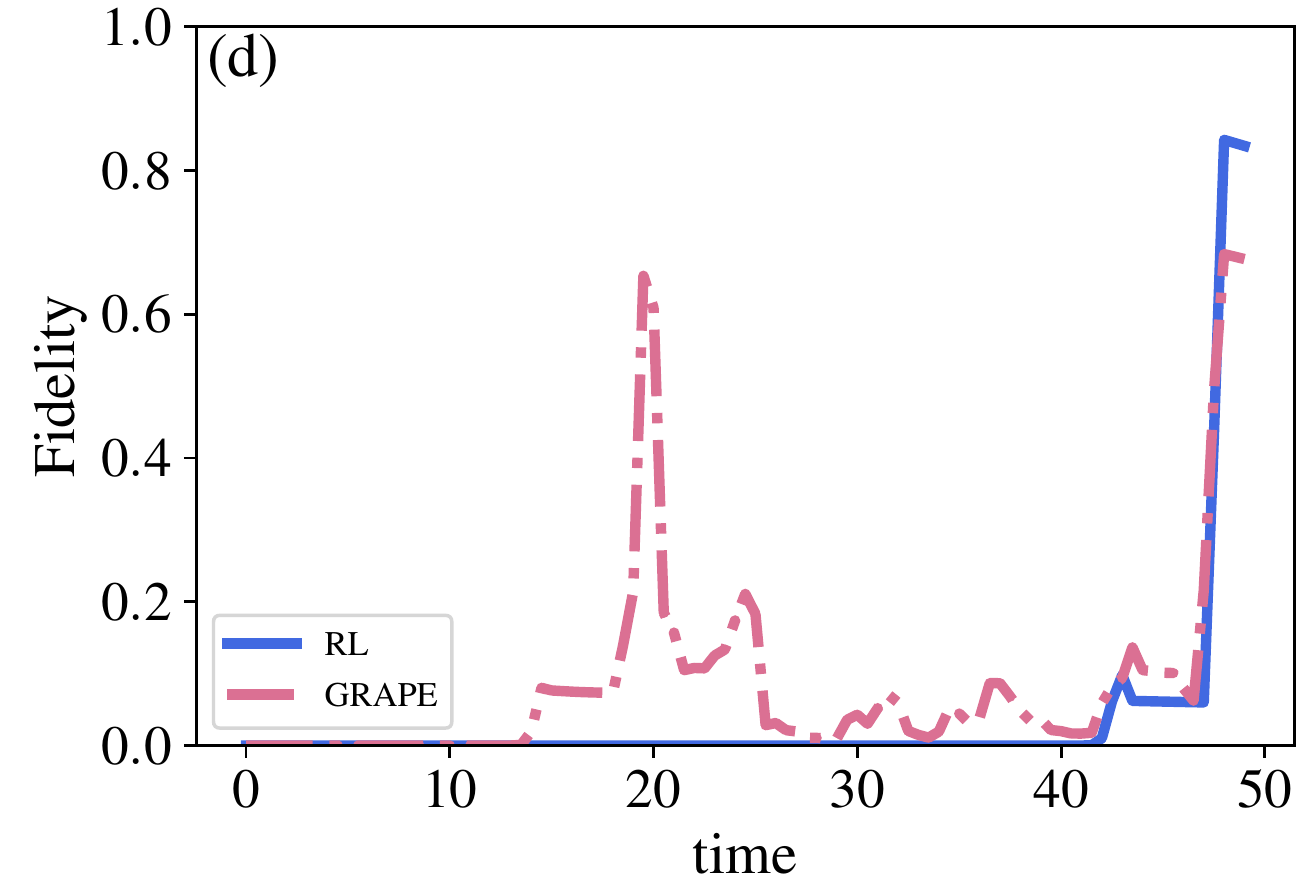}
	}
	\caption{Results from RL agent and GRAPE strategies 
		for different high level control model with 
		$\sqrt{\Gamma_l}=0.1$. The horizontal and vertical axes of 
		each subfigure denote evolution time $t$ and fidelity 
		$\mathcal{F}$.
		(A),(B),(C),(D): The evolution of fidelity with RL agent and GRAPE control 
		of
		4,6,8,10 level control model.}
	\label{fig:evol}
\end{figure}

To further understand the results shown in Fig.~\ref{fig:levd}
and~\ref{fig:levl}, we take examples from
$\sqrt{\Gamma_d},\sqrt{\Gamma_l}=0.1$ and plot the corresponding trajectories of
the fidelity in Fig.~\ref{fig:evod} and~\ref{fig:evol}. We realized
that the RL agent yields different policies according to
the types of environments: one only has to learn how to quickly
control the state to the target and decide whether to place the
control sequence at the beginning or the end of the control. As showed
in Fig.~\ref{fig:evod}, the best strategy is to fast drive the initial
state to the final state at the start of the control, since the
environment cannot change the energy of the system. While in
Fig.~\ref{fig:evod}, the strategy becomes opposed as previously shown,
the agent learns to avoid a complex control strategy to maintain the
target state but to get at the end of the control, because the energy
of the system is decaying. The trajectory of GRAPE shows there indeed have many
local minimas in the control landscape. However, the RL agent can with those those local
minimas to find optimal strategies. 

\section{Conclusion \label{con}}

We propose a quantum control framework for multi-level dissipative quantum
control optimization. The RL method is capable of finding the control
protocol that has high-fidelity of a finite dimensional quantum control
problem under disturbances and is superior to the traditional greedy
method and GRAPE algorithm. 
Moreover, RL can accommodate
switch on-off pulse shapes, which would be hard for traditional
gradient methods.

Although the control problems dealt with the different dynamics optimization
tasks, the RL agent can find high fidelity solutions with a single
set of algorithmic hyperparameters. This suggests that learning the
control landscape can be performed with minimal expert knowledge about
the physical problem.

Our results, therefore, suggest that the RL based methods can be
powerful alternatives to commonly used algorithms, capable of find
control protocols that could be more efficient in practical complex
quantum control problems. Also, the RL agent can be used to control
experimental quantum devices. The present approach is flexible enough
to be applied to different physical systems, such as qubit-cavity
systems, weak measurements, and quantum error correction. We expect
that our work would extend the deep learning techniques to deal with
more practical quantum control problems in the near future.

\begin{acknowledgements}
  This work is supported by NSF of China (Grant Nos. 11775300 and 12075310),
  NKBRSF of China (Grant No. 2014CB921202), the National Key Research and
  Development Program of China (2016YFA0300603).
\end{acknowledgements}

\appendix

\section{OPTIMAL LYAPUNOV QUANTUM CONTROL (GREEDY)
	METHOD \label{greedy}}

As the first trial, we consider a greedy way to get the optimal
strategy. Greedy algorithms are used for finding successful policies
because the algorithms are fast in converging on successful solutions
when performing local searches. To describe the greedy method more
intuitively, we use the optimal Lyapunov quantum control
theory~\cite{lyapunov1,lyapunov2,lya} to analyze the relationship
between the strength of the control field and the control fidelity.

In Lyapunov quantum control, the control fields is determined by a
Lyapunov function $f$, which will decrease with time. The evolution
of the control protocol is determined by the Eq.(\ref{eq:evo}). Further,
we assume the system satisfy the requirement for a Lyapunov function,
$f\geq 0$~\cite{qc}. The Lyapunov function can be defined as
\begin{equation}
f=\Tr(|n\rangle\langle n|\rho),
\end{equation}
The time derivative of the Lyapunov function is
given by (with $[H_0, |n\rangle\langle n|] = 0$)
\begin{equation}
\begin{aligned}
\dot{f} & = \Tr\left(|n\rangle\langle n|(-\frac{i}{\hbar}[H_0+V,\rho]+\mathcal{L}(\rho))\right) \\
& = \Tr(\mathcal{L}(\rho) |n\rangle\langle n|)-\frac{i}{\hbar} \Tr\left(\rho\left[|n\rangle\langle n|, \gamma(t) H_c\right]\right),
\end{aligned}
\end{equation}
where $\mathcal{L}(\rho)=\sum_{k} \Gamma_{k} \left(A_{k} \rho A_{k}^{\dagger}
- \frac{1}{2} \left\{A_{k}^{\dagger} A_{k}, \rho\right\} \right)$ and $H_c=\sum_{i=1}^{n-1} \left( |i\rangle  \langle i+1| + |i\rangle  \langle
i+1| \right)$.It is clear that $\dot{f}\le0$, which ensures the decreasing of the Lyapunov function. So the control function $\gamma(t)$ satisfies:
\begin{equation}
\label{eq:T}
	\Tr(\mathcal{L}(\rho) |n\rangle\langle n|)\le\gamma(t)\frac{i}{\hbar} \Tr\left(\rho\left[|n\rangle\langle n|, H_c\right]\right).
\end{equation}
Let
\begin{equation}
	\begin{array}{ll}
	C=&\Tr(\mathcal{L}(\rho) |n\rangle\langle n|)\\
	D=&\frac{i}{\hbar} \Tr\left(\rho\left[|n\rangle\langle n|, H_c\right]\right).
	\end{array}
\end{equation}

 In our problem, the control function $\gamma(t)$ is always switches between two values, so the mathematical expressions of control fields as
follows:
\begin{equation}
\gamma(t)=\left\{\begin{array}{ll}{\gamma} & {\text { if } D\geq 0, C > 0} \\ {0} & {\text { if } D\geq 0, C\leq 0} \\ {0} & {\text { if } D<0, C > 0} \\ {\gamma} & {\text { if } D<0, C\leq 0}\end{array}\right.
.
\end{equation}

\section{Two-level case without dissipative \label{2lev}}

Consider a two-level system governed by the following Hamiltonian
\begin{equation}
  H=-\frac{\omega}{2}\sigma_z+\gamma\sigma_x
\end{equation}
where we set$\hbar=1$. $\omega$is the level spacing of the
system,$\gamma=\gamma(t)$ denotes the control field. Assume that the aim is to
steer the system from an arbitrary
state$|\psi_0\rangle=\cos (\frac{\gamma_0}{2})|0\rangle+e^{i\phi}\sin(\frac{\gamma_0}{2})|1\rangle$ to
state $|1\rangle$ (target state), where $|1\rangle$ is the excited state of the
system, $|0\rangle$ is the excited state. Define a positive operator
\begin{equation}
  P_e=\mathbf{I}-|0\rangle\langle0|=|1\rangle\langle 1|
\end{equation}
The Lyapunov function can be written as
\begin{equation}
  f_e=\Tr(P_e\rho)
\end{equation}
with
\begin{equation}
  \rho=|\psi\rangle\langle \psi|,\ \ |\psi\rangle=a(t)|0\rangle+b(t)|1\rangle
\end{equation}
The Lyapunov function $f_e$ represents the overlapping between the
function $\mathbf{I}-|0\rangle\langle0|$ of target state
$|1\rangle\langle 1|$ and the actual state of the system. The time derivative of
the Lyapunov function can be calculated as follows (with
abbreviations, $a = a(t), b = b(t)$):
\begin{equation}
  \begin{aligned}
    \dot{f_e}&=\Tr(P_e\rho)=\Tr(-iP_e[-\frac{\omega}{2}\sigma_z+\gamma\sigma_x,\rho])\\
    &=\Tr(-iP_e[-\frac{\omega}{2}\sigma_z,\rho])+\Tr(-iP_e[\gamma\sigma_x,\rho])\\
    &=2\gamma \mathrm{Im}(-ab^*)
  \end{aligned}
\end{equation}
If $f_e\le0$ for all times, $f_e$ would monotonically decrease with time
under the control, meanwhile the system is asymptotically steered into
the target state $|1\rangle$. Using the method of greedy algorithm, the
control field $\gamma(t)$ takes values
\begin{equation}
  \gamma(t)=\left\{
    \begin{array}{lc}
      \gamma&(\mathrm{Im}(-ab^*)<0)\\
      0&(\mathrm{Im}(-ab^*)\ge0)\\
    \end{array}
  \right.
\end{equation}

With the optimal Lyapunov control, the time evolution of the two-level
system can be analytically calculated. In a basis spanned by
$\{|0\rangle,|1\rangle\}$, the total Hamiltonian can be expressed as
\begin{equation}
  H=\sqrt{\frac{\omega^2}{4}+\gamma^2}
  \left(
    \begin{array}{cc}
      -\cos(\theta)&\sin(\theta)\\
      \sin(\theta)&\cos(\theta)
    \end{array}
  \right)
\end{equation}
with $\theta$ defined by
\begin{equation*}
  \tan\theta=\frac{2f}{\omega}
\end{equation*}
The eigenvalues of the Hamiltonian H are
\begin{equation}
  E_\pm=\pm\sqrt{\frac{\omega^2}{4}+\gamma^2}
\end{equation}
and the corresponding eigenvectors are given by,
\begin{equation*}
  |E_+\rangle=-\cos\frac{\theta}{2}|0\rangle+\sin\frac{\theta}{2}|1\rangle
\end{equation*}
\begin{equation*}
  |E_-\rangle=\sin\frac{\theta}{2}|0\rangle+\cos\frac{\theta}{2}|1\rangle
\end{equation*}
The time evolution operator can be calculated to be
\begin{widetext}
  \begin{eqnarray}
    U=exp(-iHt)=\left(\begin{array}{cc}
                        e^{-iE_-t}\cos^2\frac{\theta}{2}+e^{-iE_+t}\sin^2\frac{\theta}{2}&\frac{1}{2}(e^{-iE_+t}-e^{-iE_-t})\sin\theta\\
                        \frac{1}{2}(e^{-iE_+t}-e^{-iE_-t})\sin\theta&e^{-iE_-t}\sin^2\frac{\theta}{2}+e^{-iE_+t}\cos^2\frac{\theta}{2}
                      \end{array}\right)
  \end{eqnarray}
\end{widetext}
In the absence of a control field (i.e.,$\gamma(t)=0$), we have
$\theta=0$. The time evolution operator reduces to a diagonal form,
\begin{equation*}
  U=\left(
    \begin{array}{cc}
      e^{\frac{i\omega t}{2}}&0\\
      0&e^{\frac{-i\omega t}{2}}
    \end{array}
  \right)
\end{equation*}

Assume that the initial state of a two-level system is
\begin{equation*}
  |\psi_0\rangle=\cos (\frac{\gamma_0}{2})|0\rangle+e^{i\phi}\sin(\frac{\gamma_0}{2})|1\rangle=a_0|0\rangle+b_0|1\rangle
\end{equation*}
With different parameters $\gamma_0$and $\psi$, $|\psi _0\rangle$ can represent an
arbitrary pure state. Let the target state $|1\rangle$ correspond to the
south pole on the Bloch sphere. Since
$\mathrm{Im}(-a_0b_0^*)=-\frac{\sin\phi\sin\gamma_0}{2}$, the first control
field is calculated as,
\begin{equation}
  \gamma(t)=\left\{
    \begin{array}{lc}
      \gamma&(\mathrm{Im}(-ab^*)<0),(0<\theta<\pi)\\
      0&(\mathrm{Im}(-ab^*)\ge0),(\pi\le\theta<2\pi, \theta=0)\\
    \end{array}
  \right.
\end{equation}
Assume that this control would last until time $\tau$ ; i.e., the
duration of this control is $\tau$ . With this control, the state evolves
to

\begin{widetext}
  \begin{eqnarray}
    \begin{aligned}
      |\psi_\tau\rangle = &
      [(e^{-iE_-t}\cos^2\frac{\theta}{2}+e^{-iE_+t}\sin^2\frac{\theta}{2})
      \cos\frac{\gamma_0}{2}+\frac{1}{2}(e^{-iE_+t}-e^{-iE_-t}) \sin\theta
      e^{i\phi}\sin\frac{\gamma_0}{2}]|0\rangle\\
      & + [\frac{1}{2} (e^{-iE_+t}-e^{-iE_-t}) \sin\theta\cos\frac{\gamma_0}{2}
      + (e^{-iE_-t}\sin^2\frac{\theta}{2}+e^{-iE_+t}\cos^2\frac{\theta}{2})
      e^{i\phi}\sin\frac{\gamma_0}{2}]|1\rangle\\
      & \equiv a_\tau|0\rangle+b_\tau|1\rangle
    \end{aligned}
  \end{eqnarray}
\end{widetext}
From the design of the control law , we find that a control field
would last until $\mathrm{Im}(-a_\tau b_tau^*)$ changes sign. Then
$\tau$ can be given by solving $\mathrm{Im}(-a_\tau b_tau^*)=0$: Meanwhile,
the sign of $\mathrm{Im}(-a_\tau b_tau^*)$ determines the next control
field. Simple algebra shows that
\begin{widetext}
  \begin{equation}
    \mathrm{Im}(-a_\tau b_\tau^*) = \frac{1}{2} ( \sin (2E_-\tau) (\cos \theta \sin
    \gamma_0 \cos \phi+\sin \theta \cos \gamma_0)+\sin \gamma_0 \sin \phi \cos (2E_-\tau)) 
  \end{equation}
\end{widetext}

\section{Distributed Proximal Policy Optimization \label{PPO}}

The Actor-Critic algorithm combines the advantages of policy-based and
value-based methods. While the PPO algorithm~\cite{DPPO1,DPPO2} based
on Actor-Critic aims to optimize policy update. The central idea of
Proximal Policy Optimization is to avoid having too large policy
update which is proposed by trust region policy optimization
(TRPO)~\cite{TRPO}. The underlying idea of such improvements thereby
is limiting the magnitude of updates to $\theta$ by imposing constraints on
the difference between $\pi_{\theta_{\text { old }}}$ and $\pi_\theta$ in order to
prevent catastrophic jumps out of optima and achieve a better
convergence behavior.

\begin{figure}[htbp]
  \includegraphics[width=0.98\columnwidth{}]{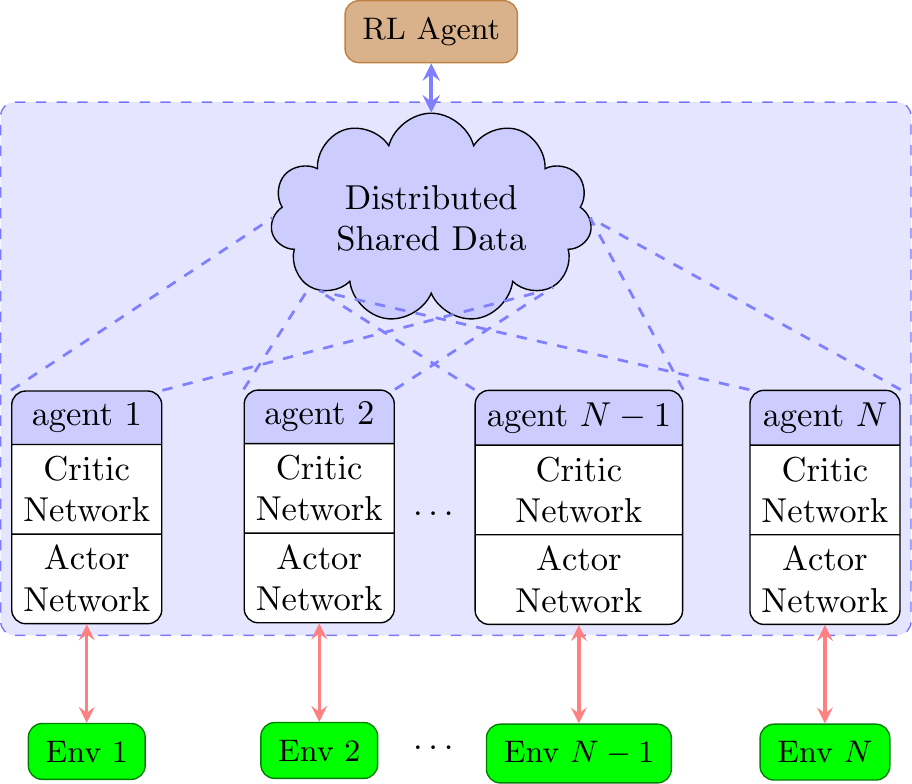}
  \caption{\label{dppo} Schematics of the DPPO algorithm. Data
    collection and gradient calculation are distributed over workers,
    labeled as "agent i". Then the weights of RL agent update
    synchronously. The environments, labeled as “env i”. }
\end{figure}

One main novelty hereby lies in the introduced loss of DPPO,
\begin{equation}
  \begin{aligned}
    L^{C L I P}(\theta) = & \mathbb{E}_{t} \left[\min \left(r_{t}(\theta)
        A_t(s,a)\right.\right.,\\
    & \left.\left.\operatorname{clip}\left(r_{t}(\theta), 1-\epsilon, 1+\epsilon\right)
        A_{t}(s,a)\right)\right]
  \end{aligned}
  \label{eq:ppo}
\end{equation}
where $\mathbb{E}_{t}$ and $A_{t}(s,a)$ are the expectation over time
steps and the advantage at time t respectively.  If
$r_{t}(\theta)>1$, the action is more probable in the current policy than
the old policy; if $r_{t}(\theta)>1$ is between 0 and 1, the action is less
probable for current policy than for the old one.

As consequence, a new objective function from Eq.(\ref{eq:pg}) could
be
\begin{equation}
  L^{C P I}(\theta) = \hat{\mathbb{E}}_{t} \left[\frac{\pi_{\theta}\left(a |
  	s\right)}{\pi_{\theta_{\text { old }}}\left(a | s\right)}
    A_{t}(s,a)\right] = \hat{\mathbb{E}}_{t}\left[r_{t}(\theta)
    A_{t}(s,a)\right]. 
\end{equation}
However, without a constraint, if the action taken is much more
probable in our current policy than in our former, this would lead to
a large policy gradient step and consequence an excessive policy
update.

So the PPO algorithm clip probability ratio directly in the objective
function with its Clipped surrogate objective
function[Eq.~\ref{eq:ppo}]. The loss function poses a lower bound on
the improvement induced by an update and hence establishes a trust
region around $\pi_{\theta_{\text { old }}}$. The hyperparameter
$\theta$ controls the maximal improvement and thus the size of the trust
region.

\begin{algorithm}[htbp]
  \caption{\centering Distributed Proximal Policy Optimization}
  \label{DPPO}
  Randomly initialize critic network $V_\phi(s)$ and actor
  $\pi_{\theta}(a|s)$ with weights $\phi$ and $\theta$;\
	
  \For{$iteration \in\{1,2,\dots,C\}$}
  {
    \For{$actor=0, \dots, N$}
    {Initialize $s_{0}$;\
			
      Run policy $\pi_{\theta}$ $\frac{T}{\delta t}$ times, collecting
      $\{s_{t},a_{t},R_{t+1}\}$;\
			
      Estimate advantages
      $A_{t}=\sum_{t^{\prime}>t} \gamma^{t^{\prime}-t}
      R_{t^{\prime}}-V_{\phi}\left(s_{t}\right)$;\
      
      Estimate $\hat{V}_{t}=A_{t}+V_{\phi}\left(s_{t}\right)$;\ 
    }

    $\pi_{\theta_{\mathrm{old}}} \leftarrow \pi_{\theta}$
			
    \For{$j \in\{1,\dots,M\}$}
    {
      $J_{PPO}(\theta)= \left[\min \left(r_{t}(\theta)
      A_t\right.\right.,
       \left.\left.\text{clip}\left(r_{t}(\theta), 1-\epsilon, 1+\epsilon\right)
      A_{t}\right)\right]$;\
				
      Update $\theta$ by a gradient method w.r.t. $J_{P P O}$;\
      $J_{critic}(\phi) = -\mathbf{E}_{t}\left(\hat{V}_{t} -
        V_\phi\left(s_{t}\right) \right)^{2}$;\
      
      Update $\phi$ by a gradient method w.r.t. $J_{\text{critic }}(\phi)$
    }
  }
\end{algorithm}

In order to improve the efficiency of the learning process, a
distributed version of PPO algorithm (DPPO)~\cite{DPPO1}, is
implemented in our calculation[Fig.~\ref{dppo}].Algorithm~\ref{DPPO}
shows the pseudocode for the DPPO.

\section{The effect of distribution of eigenenergies \label{eng}}

In the main text we use a regular distribution of eigeneneriges to test our algorithm. However, a different distribution of eigenenergy would affect the performance of the algorithm. The effect of distribution of eigenenergies was examined for two example cases, (\romannumeral1) the eigenenergy $E_i$ extracted from the uniform distribution with $E_1=0.40252154, E_2=0.68846289, E_3=0.8557115, E_4=0.25471114$ and (\romannumeral2) the eigenenergy $E_i$ has degenerate in the middle with
$E_1=1, E_2=E_3=2, E_4=3$.

\begin{figure}
	\centering
	\subfloat{
		\includegraphics[width=0.24\textwidth]{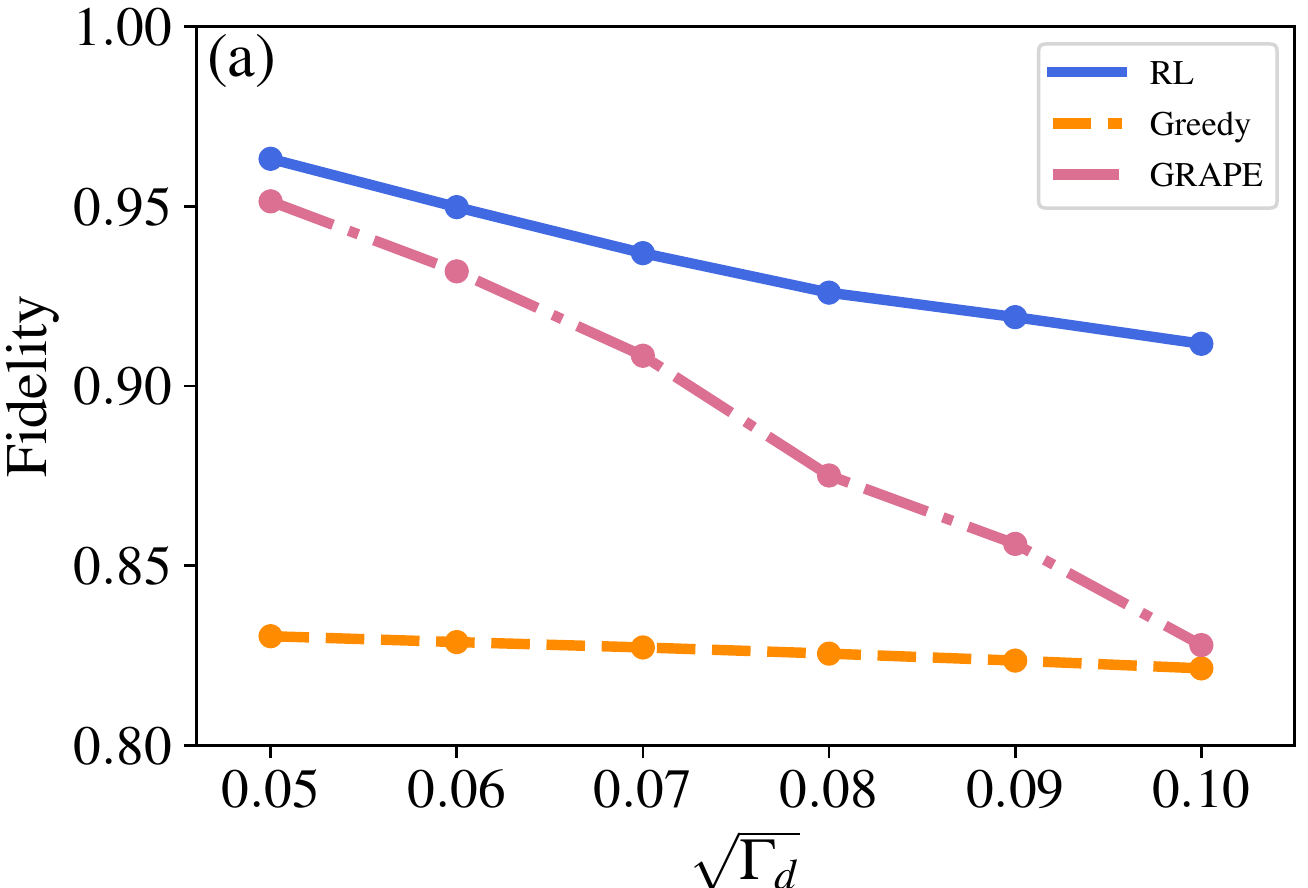}
	}
	\subfloat{
		\includegraphics[width=0.24\textwidth]{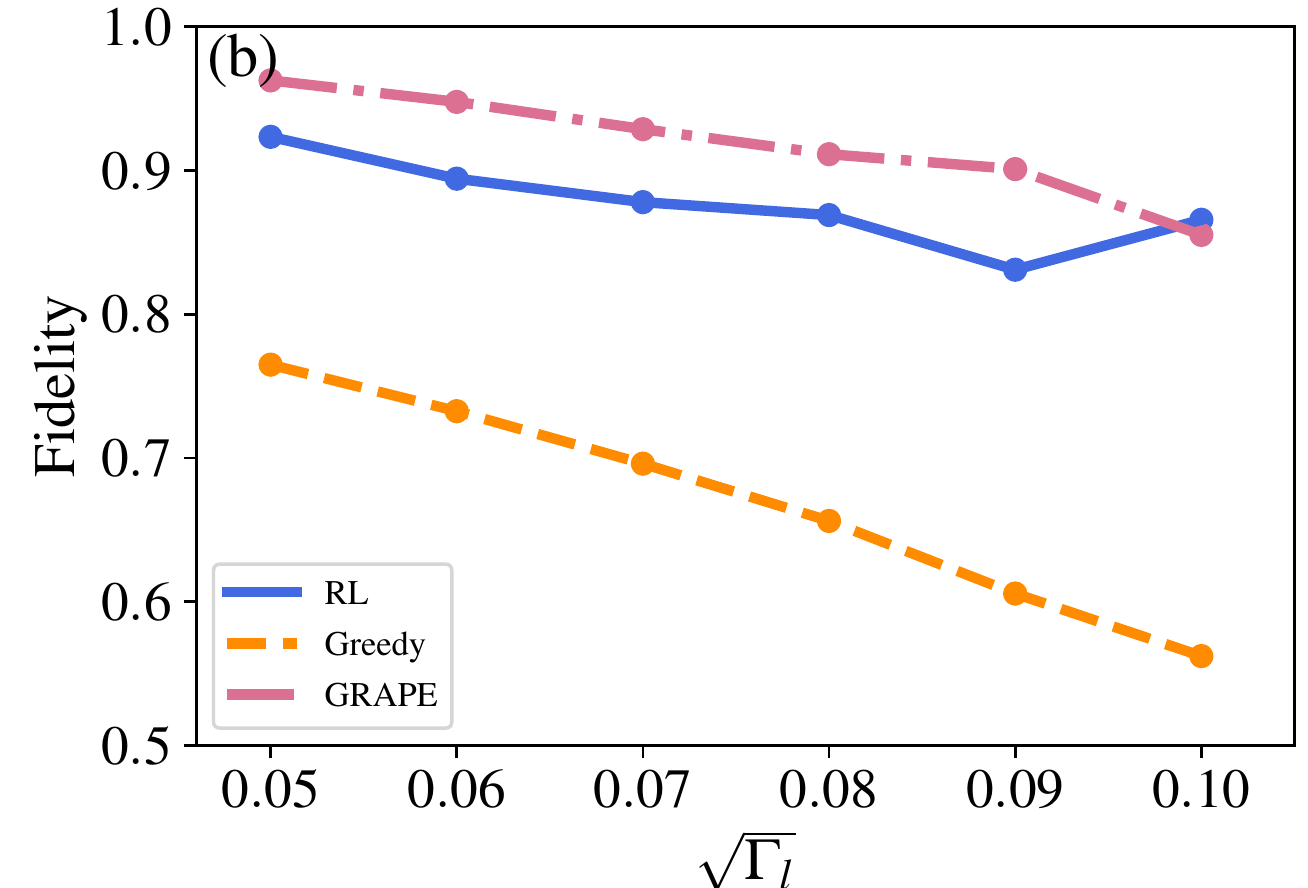}
	}

	\caption{Results from the three algorithms
		for 4 level control model with the Hamiltonian of (\romannumeral1) under disturbances.  The horizontal and vertical axes of  each 
		subfigure denote noise rate $\sqrt{\Gamma_{k,n}}$ and 
		fidelity $\mathcal{F}$. 
		(A): dephasing dynamics (B): energy decay dynamics.
}
	\label{fig:ran}
\end{figure}

\begin{figure}
	\centering
	\subfloat{
		\includegraphics[width=0.24\textwidth]{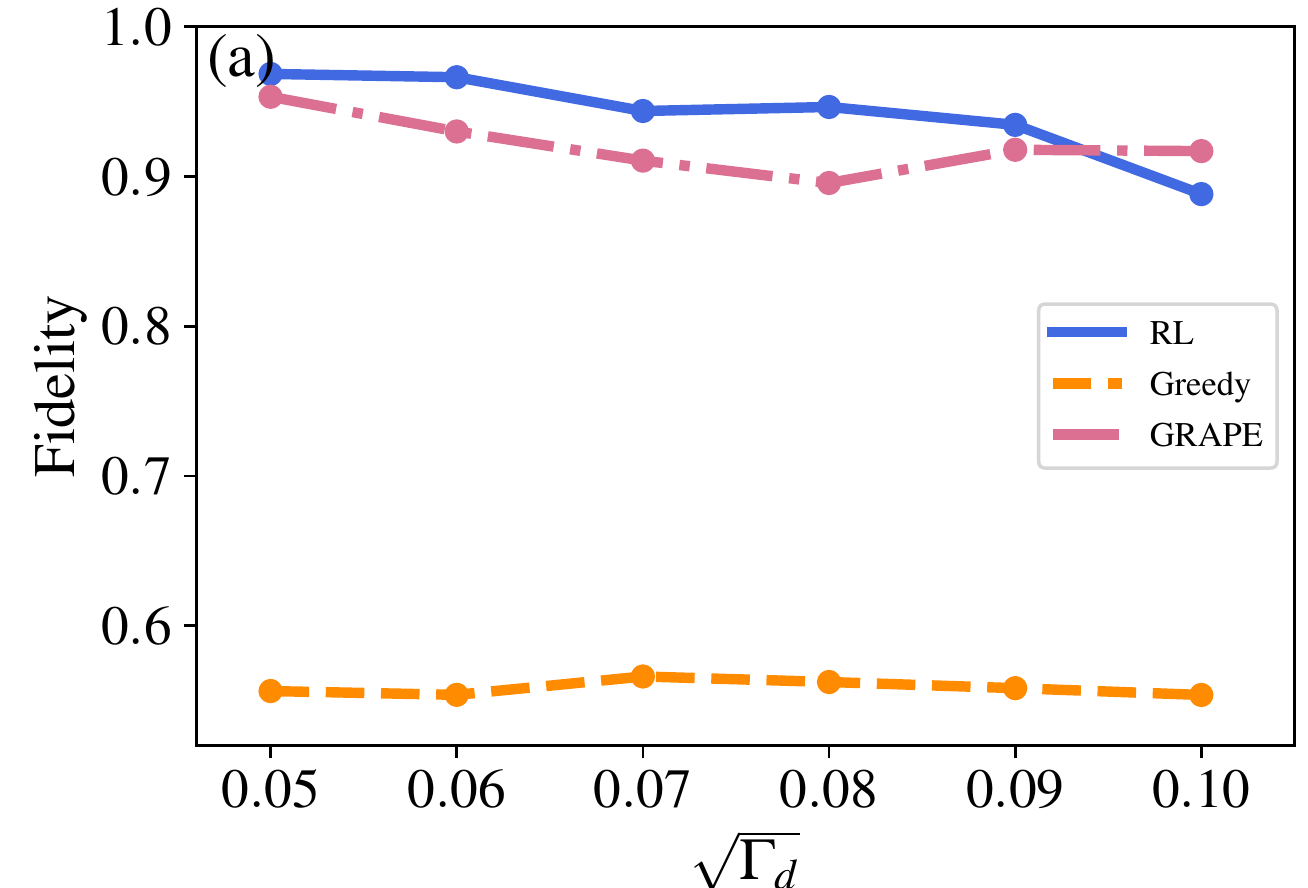}
	}
	\subfloat{
		\includegraphics[width=0.24\textwidth]{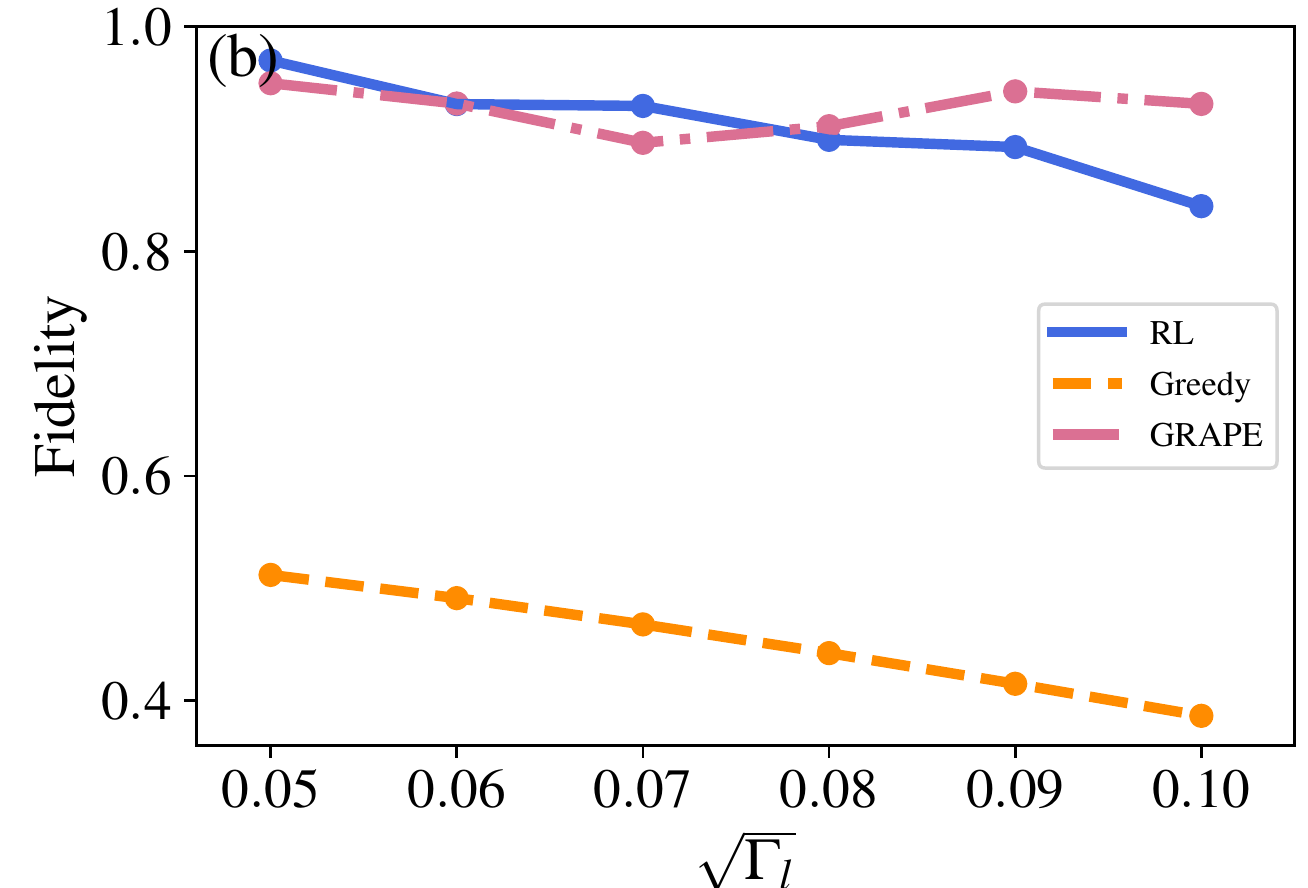}
	}

	\caption{Results from the three algorithms
		for 4 level control model with the Hamiltonian of (\romannumeral2) under disturbances.  The horizontal and vertical axes of  each 
		subfigure denote noise rate $\sqrt{\Gamma_{k,n}}$ and 
		fidelity $\mathcal{F}$. 
		(A): dephasing dynamics (B): energy decay dynamics. }
	\label{fig:deg}
\end{figure}

As Fig.~\ref{fig:ran} and Fig.~\ref{fig:deg} shown, the performance of the three algorithms is affected under the different energy distribution. However, the GRAPE algorithm and our algorithm still maintain superiority over the greedy algorithm. This is consistent with what we discussed in the main text.

\section{Hyper-Parameters and Learning Curves}

Our RL agent makes use of two deep neural networks to approximate the
values for the possible actions of each state and the optimal policy.
Each network consists of 4 layers. All layers have ReLU activation
functions except the output layer which has linear activation. The
hyper-parameters of the network are summarized in Table~\ref{para}.

All algorithms are implemented with Python 3.6, and have been run on
two 14-core 2.60GHz CPU with 188 GB memory and four GPUs.

\begin{table}
  \begin{threeparttable}
    \caption{Training Hyper-Parameters}
    \begin{tabular}{cc}
      \textrm{Hyper-parameter} &\textrm{Values} \\
      \hline
      Neurons in actor network & $\{1024,1024,1024,1024\}$\\
			
      Neurons in critic network & $\{1024,1024,1024,1024\}$\\
			
      Actor numbers N & 12\\
			
      Batch size & \tnote{a} \\
			
      PPO clipping $\epsilon$ & 0.2\\
			
      Learning rate & $0.0001$\tnote{b}\\
			
      Update steps M & 15\\
			
      Reward decay $\Gamma$ & 0.85\\
			
      Total episode C
      
       & \tnote{c} \\
			
      \label{para}
    \end{tabular}
    \begin{tablenotes}\footnotesize
    \item [a] is the same as the time steps
    \item[b] With Adam algorithm
    \item[c] different for various tasks
    \end{tablenotes}
  \end{threeparttable}
\end{table}

\FloatBarrier

\bibliographystyle{apsrev4-2} \bibliography{quantum_ladder.bbl}

\end{document}